\documentclass[11pt,tightenlines,eqsecnum,floats,aps,amssymb,nofootinbib,prd,shownopacs,floatfix,superscriptaddress]{revtex4-2}
\usepackage{graphicx}
\usepackage{epstopdf}
\usepackage{latexsym}
\usepackage{amssymb}
\usepackage{amsmath}
\usepackage{color}
\usepackage{mathrsfs}
\usepackage{xparse}
\usepackage[dvipsnames]{xcolor}
\usepackage{float}
\usepackage{dsfont}
\usepackage{mathtools}
\usepackage[toc,title,page]{appendix}
\usepackage[colorlinks=false, pdfborder={0 0 0}]{hyperref}

\usepackage{tensor}
\usepackage{braket}

\numberwithin{equation}{section}

\begin{document}
\newpage
\begin{abstract}
We investigate gravitational dust collapse within an effective loop quantum gravity (LQG)–inspired model exhibiting an asymmetric bounce in the marginally bound case. This work extends previous studies, which have predominantly focused on models with either symmetric bounces or asymmetric bounces restricted to homogeneous dust configurations. Our analysis emphasises the phenomenological implications of the model through a combination of analytical and numerical investigations, with particular attention to singularity resolution and the formation of trapped surfaces. As in symmetric bounce models, the central curvature singularity inside the collapsing dust cloud is resolved. However, in contrast to the symmetric case, we find that a singularity emerges in the polymerised vacuum region during the bounce phase. This singularity can be identified as a shell-crossing singularity and exhibits the expected power-law behaviour of curvature scalars. Furthermore, likewise to the symmetric bounce models, we find a critical mass threshold governing the formation of inner and outer horizons in the pre-bounce phase. No analogous critical mass restriction arises for the formation of the inner horizon in the post-bounce phase, highlighting a qualitative difference between the pre- and post-bounce dynamics.

\end{abstract}
\title{Investigation of the gravitational dust collapse of the LQG-inspired effective asymmetric bounce model}

\author{Kristina Giesel}
\email{kristina.giesel@fau.de}
\affiliation{Institute for Quantum Gravity, Theoretical Physics III, Department of Physics,  Friedrich-Alexander-Universit\"at Erlangen-N\"urnberg, Staudtstr. 7, 91058 Erlangen, Germany.}

\author{Hongguang Liu}
\email{liuhongguang@westlake.edu.cn}
\affiliation{Institute for Theoretical Sciences and Department of Physics, Westlake University, Hangzhou 310024, Zhejiang, China}

\author{Eric Rullit}
\email{eric.rullit@fau.de}
\affiliation{Institute for Quantum Gravity, Theoretical Physics III, Department of Physics,  Friedrich-Alexander-Universit\"at Erlangen-N\"urnberg, Staudtstr. 7, 91058 Erlangen, Germany.}

\maketitle
\section{Introduction}
\label{sec:Intro}
The investigation of gravitational dust collapse within loop quantum gravity (LQG)–inspired effective models has attracted growing interest as a setting in which to explore how quantum gravity corrections alter the fate of classical singularities. 
In spherically symmetric models, such corrections to classical general relativity are commonly implemented through polymerisation functions \cite{Ashtekar:2005qt,Modesto:2005zm,Boehmer:2007ket,Chiou:2012pg,Gambini:2013hna,Brahma:2014gca,Dadhich:2015ora,Tibrewala:2013kba,BenAchour:2017ivq,Yonika:2017qgo,DAmbrosio:2020mut,Olmedo:2017lvt,Ashtekar:2018lag,Ashtekar:2018cay,Bojowald:2018xxu,BenAchour:2018khr,Bodendorfer:2019cyv,Alesci:2019pbs,Assanioussi:2019twp,Benitez:2020szx,Gan:2020dkb,Gambini:2020qhx,Husain:2021ojz,Husain:2022gwp,Li:2021snn,Gan:2022mle,Kelly:2020uwj,Gambini:2020nsf,Han:2020uhb,Zhang:2021xoa,Munch:2022teq,Lewandowski:2022zce,Giesel:2021dug,Giesel:2022rxi,Han:2022rsx,Giesel:2023tsj,Giesel:2024mps,Cafaro:2024vrw}. For LQG-inspired constructions, these polymerisation functions are typically chosen to be bounded, reflecting the underlying quantization in terms of holonomies within LQG. In addition, polymerisation functions can incorporate inverse-triad corrections, thereby encoding effects associated with inverse volume contributions known from the full LQG. Allowing for a broader class of polymerisation functions—including unbounded ones—extends this framework beyond strictly LQG-inspired modifications. In this language, other effective geometries in spherical symmetry, such as regular black hole solutions of Bardeen or Hayward type, can be described within the same formalism \cite{Giesel:2024mps}. A key advantage of introducing polymerisation functions is therefore that qualitative properties and physical implications can often be analysed for entire classes of models, rather than requiring a case-by-case study. Furthermore, a relation to four-dimensional covariant actions in the context of extended mimetic gravity models has been established allowing to embed these spherically symmetric effective models into the field of modified gravity \cite{Giesel:2024mps,Giesel:2025kdl}. 
~\\
~\\
In this work, we limit ourselves to LQG-inspired polymerisations. In addition we consider effective spherically symmetric models that allow a consistent reduction to its LTB sector and for which the energy density is conserved. As a further requirement we want to restrict to those classes of models for which an extended mimetic model as an underlying covariant Lagrangian exist. The latter requires that no inverse triad corrections are present \cite{Giesel:2024mps}. The most widely studied example in this context is the effective model of gravitational dust collapse derived from the symmetric bounce scenario of loop quantum cosmology \cite{Bojowald:2008ja,Bojowald:2009ih,Munch:2020czs,Husain:2021ojz,Giesel:2021dug,Giesel:2022rxi,Lewandowski:2022zce,Fazzini:2023scu,Han:2023wxg,Fazzini:2023ova,Cipriani:2024nhx,Han:2024ydv}. In the spherically symmetric, marginally bound LTB setting, this construction has the remarkable property that the dynamics decouples completely along the radial direction: the Hamiltonian density at each radial point is governed by the same effective dynamics that underlies the symmetric LQC bounce. For homogeneous dust profiles, analyses of the effective Oppenheimer–Snyder collapse based on this model can be found in \cite{Ou:2025bbv}, where it is shown that the classical central singularity is resolved. However, when inhomogeneous dust profiles are considered, it has been demonstrated that in this effective LTB model with a symmetric bounce, shell-crossing singularities generically arise \cite{Fazzini:2023ova}.
~\\
~\\
At the same time with the presence of shell-crossing singularities, recent analyses have emphasised the need for suitable spacetime extensions. Approaches based on weak solutions \cite{Lasky:2006hq,Husain:2022gwp,Fazzini:2025zrq}, shock-wave analogies \cite{Liu:2025fil}, or the treatment of isolated thin shells as proxies for post–shell-crossing configurations \cite{Sahlmann:2025fde,Fazzini:2025nse} highlight that the physically relevant issues are not limited to the resolution of central curvature singularities alone.
~\\
~\\
Far less studied in connection with effective marginally bound LTB models is the case of an asymmetric bounce, which is also known from LQC, when a Thiemann regularisation is implemented prior to symmetry reduction and, in particular, the Lorentzian part of the Hamilton constraint is retained \cite{Yang:2009fp}. This construction was further developed and analysed in \cite{Li:2018opr}. An alternative method leading to the same effective dynamics was presented in \cite{Dapor:2017rwv}, where a non-graph-changing quantisation within  full LQG was considered and the effective dynamics were derived by calculating expectation values with respect to semiclassical cosmological states. Similar result can be obtained from the coherent state path integral formulation of the full LQG \cite{Han:2019vpw} and the $\bar{\mu}$-scheme dynamics can be obtained from a dynamical lattice or extended phase space \cite{Han:2019feb,Han:2021cwb,Giesel:2023euq}.
A recent investigation on an effective Oppenheimer-Snyder model with an asymmetric bounce can be found in \cite{Ou:2025bbv} in which the main focus lies on the quasi-normal modes and thermodynamic properties of the black hole. 

In contrast, the present work extends the results of \cite{Giesel:2023hys}, where first steps toward the analysis of the marginally bound effective LTB model with an asymmetric bounce were undertaken, including the derivation of an analytical parametric solution and the computation of curvature scalars. Here, we pursue a more detailed investigation of the model, focusing on the phenomenology of the effective dust collapse process, with particular emphasis on the formation of trapped surfaces and the behaviour in the vicinity of shell-crossing singularities. This allows us to go beyond merely establishing the singularity resolution and gain new insights into the physical implications of the model, with the analysis extending beyond the effective Oppenheimer–Snyder model considered in \cite{Ou:2025bbv}.

The paper is structured as follows: In section \ref{sec:Intro} we provide the introduction and outline the motivation and scope of the work. Section \ref{sec:EffModel} reviews effective spherically symmetric models with generalised polymerisations and establishes the framework used throughout the analysis of the effective gravitational dust collapse model with asymmetric bounce in  section \ref{sec:AsymB}. In section \ref{sec:AsymB} we start with introducing the effective asymmetric bounce model  and first discuss its effective dynamics in subsection \ref{sec:IIIED}, followed by an analysis of its phenomenological implications in subsection \ref{sec:IIIPI}. This includes a study of strong and weak singularities in subsection and the formation of (anti-)trapped surfaces, and numerical results for inhomogeneous dust collapse that underline the analytical results discussed before in subsection \ref{sec:IIINR}, as well as a comparison with the symmetric bounce scenario in subsection \ref{sec:IIISB}. We conclude in section \ref{sec:Concl} with a summary and outlook. An appendix collects additional material on curvature invariants relevant for the discussion and results in the main text.

\section{Effective spherically symmetric models with generalised polymerisations}
\label{sec:EffModel}
In this section we will briefly introduce the formulation of spherically symmetric models within the framework of effective LQG-inspired models, which will be needed for the later investigation of a specific model that involves an asymmetric bounce. We will start with a brief review of the classical connection formulation of spherically symmetric models with dust together with the classical reduction to LTB spacetimes via the so-called LTB condition. Following closely \cite{Giesel:2023tsj, Giesel:2024mps} we will then sketch the main ideas of lifting the classical formalism to the effective description by implementing quantum modifications in terms of generalised polymerisation functions.
~\\
~\\
The class of models considered in this work are subject to symmetry reduction of the classical phase space in GR coordinatised by the Ashtekar-Barbero variables $(A_a^j, E_j^a)$, where $A_a^j$ denotes an SU(2) connection and $E_j^a$ the canonically conjugate densitised triad. In the case of spherical symmetry the topology of the spatial manifold is of the form $\sigma \cong \mathbb{R} \times S^2$, for which the Ashtekar-Barbero connection and the densitised triad after fixing the Gau\ss{} constraint read \cite{Bojowald:2008ja,Bojowald:2009ih}
\begin{align}
    A_a^j \tau_j dX^a &= 2 \gamma K_x(x) \tau_1 dx + \left(\gamma K_\phi(x) \tau_2 + \frac{\partial_x E^x(x)}{2 E^\phi(x)} \tau_3\right) d\theta \nonumber \\
    &+\left(\gamma K_\phi(x) \tau_3 - \frac{\partial_x E^x(x)}{2 E^\phi(x)} \tau_2\right) \sin (\theta) d\phi + \cos (\theta) \tau_1 d\phi \\
    E_j^a \tau^j \frac{\partial}{\partial X^a} &= E^x(x) \sin(\theta) \tau_1 \partial_x + E^\phi(x) \sin(\theta) \tau_2 \partial_\theta + E^\phi(x) \tau_3 \partial_\phi.
\end{align}
Here $X^a = (x, \theta, \phi)$ denote the (spatial) spherical coordinates, $\gamma$ is the Barbero-Immirzi parameter and $\tau_i := -\sigma_j/2$ where $\sigma_j$ denote the Pauli matrices. In a non-rotational dust model, we consider an additional canonical dust field and its conjugate momentum $(T, P_T)$ and using a relational perspective we can choose the dust field as the dynamical reference field for the temporal evolution. Then the proper time measured along the radial dust word lines provides a physical temporal coordinate that can be implemented by using the dust time gauge $T - t = 0$. In the case of spherical symmetry this leads to a partially gauge fixed phase space with the following non-vanishing Poisson brackets \cite{Giesel:2023tsj}
\begin{align}
    \{K_x(x), E^x(y)\} = G \delta(x, y),\hspace{.5cm}\{K_\phi(x), E^\phi(y)\} = G \delta(x, y),\hspace{.5cm}\{T(x), P_T(y)\} = \delta(x, y).
\end{align}
Thus, after defining the set $(t, x, \theta, \phi)$ of comoving coordinates, the general spherically symmetric line element in terms of components of the densitised triad can be written as
\begin{align}\label{eq:LE1}
    ds^2 = -dt^2 + \frac{(E^\phi)^2}{|E^x|}(dx + N^x dt)^2 + |E^x|(d\theta^2 + \sin^2(\theta) d\phi^2),
\end{align}
where $N^x$ denotes the radial component of the shift vector. The dynamics of the gravitational sector is governed by the partially gauge-fixed Hamiltonian
\begin{align}
    H[N^x] = \int_\sigma dx (C + N^x C_x)(x),
\end{align}
where $C$ and $C^x$ denote the gravitational parts of the Hamiltonian and spatial diffeomorphism constraint respectively given by
\begin{align}
    C(x) &= \frac{1}{2G}\frac{E^\phi}{\sqrt{E^x}} \left[-E^x\left(\frac{4 K_x K_\phi}{E^\phi} + \frac{K_\phi^2}{E^x}\right) + \left(\frac{E^{x\prime}}{2 E^\phi}\right)^2 - 1 + 2 \frac{E^x}{E^\phi}\left(\frac{E^{x\prime}}{2 E^\phi}\right)^\prime\right](x) \label{eq:ClassHam}\\
    C_x(x) &= \frac{1}{G} (E^\phi K_\phi^\prime - K_x E^{x\prime})(x),
\end{align}
where the prime denotes derivatives with respect to the radial coordinate $x$. In this work we will restrict our choice to non-rotational dust and since it will serve as a dynamical reference frame for the Hamiltonian constraint, the details of its contributions to the spatial diffeomorphism and Hamiltonian constraint are not relevant for the dynamics on the partially reduced phase space that we consider here in which the Hamiltonian constraint deparametrises. For a comprehensive discussion on both the classical and effective formulation, also beyond the non-rotational dust model, the reader is referred to \cite{Giesel:2023tsj}.
~\\
~\\
As a specific solution to the spherically symmetric Einstein's field equations we now consider LTB spacetimes describing the gravitational collapse of (inhomogeneous) pressureless dust spheres. The LTB line element in terms of comoving coordinates from above reads \cite{Lemaitre:1933gd, Tolman:1934za, Bondi:1947fta}
\begin{align}
    ds^2 = -dt^2 + \frac{(R^\prime)^2}{1 + \mathcal{E}(x)} dx^2 + R^2 (d\theta^2 + \sin^2(\theta) d\phi^2),
\end{align}
where $R(t, x)$ is the areal radius of the dust shell located at the radial coordinate $x$ and $\mathcal{E}(x)$ its total energy. The areal radius $R(t, x)$ is subject to a Friedmann equation in which $\mathcal{E}(x)$ enters as one of the source terms
\begin{align}
\label{eq:ClassFRW}
\Dot{R}^2 = \mathcal{E}(x) + \frac{2 G M(x)}{R}
\end{align}
following from Einstein's field equations which includes the Misner-Sharp mass $M(x)$ measuring the total gravitational mass confined inside the shell $x$. To make a closer contact to the above canonical formulation we define $R(t,x) := \sqrt{E^x(x)}$ and notice that the above LTB metric can be obtained from \eqref{eq:LE1} by setting $N^x = 0$ and enforcing the so-called LTB condition \cite{Bojowald:2008ja,Bojowald:2009ih}
\begin{align}
    C_{\mathrm{LTB}}(x) := (|E^x|^\prime - 2 E^\phi \sqrt{1 + \mathcal{E}(x)})(x) = 0.
\end{align}
The one physical degree of freedom in spherically symmetric dust models is encoded in two canonically conjugate phase space functions, that after completing fixing the gauge in LTB models can be chosen to be the canonical pair $(K_\phi, E^x)$ as outlined in \cite{Giesel:2023tsj} in section 2.1 and 2.2 via a detailed constraint analysis. This pair satisfies the following evolution equations
\begin{align}\label{eq:EOM1}
    \partial_t E^x = -2 K_\phi \sqrt{E^x},\hspace{.5cm}\partial_t K_\phi = \frac{1}{2 \sqrt{E^x}} (K_\phi^2 + \mathcal{E}(x)).
\end{align}
The gravitational part of the LTB Hamiltonian constraint can be written as
\begin{align}
    C(x)|_{\mathrm{LTB}} = \frac{\partial_x \widetilde{H}(x)}{\sqrt{1+\mathcal{E}(x)}},\hspace{.5cm}\widetilde{H}(x) = -\frac{1}{2G}\left(\sqrt{E^x}K_\phi^2-2\mathcal{E}(x)\right) = -M(x).
\end{align}
In the  marginally bound LTB model where $\mathcal{E}(x) = 0$ the system is characterised by the property that the kinetic energy of each dust shell is compensated by the same amount of gravitational potential energy. By virtue of the dynamical equations in \eqref{eq:EOM1} the decoupled Hamiltonian densities $H(x)$ of a fixed shell at $x$ are in fact time independent and moreover - up to a minus sign - equal to the conserved quantity $M(x)$.
The Friedmann equation in \eqref{eq:ClassFRW} can then be easily recovered by writing the radial component of the densitised triad in terms of the areal radius of the dust shell.
~\\
~\\
Classical LTB models are characterised by a central singularity at $R=0$. In addition, weak singularities known as shell-crossing singularities may occur. Looking at the form of the dust energy density $\rho$, which is given by
\begin{align}
\rho(t, x) = \frac{M^\prime(x)}{4 \pi R(t, x)^2 R^\prime(t, x)}
\end{align}
we see that it diverges for $R^\prime=0$ if, in addition, $M^\prime\not=0$, which is precisely the criterion for the occurrence of shell-crossing singularities.
~\\
~\\
After defining spherically symmetric models with dust in the Hamiltonian formulation of classical GR we now proceed to their effective counterparts. These so-called effective models carry quantum geometric corrections modifying  (part of) the classical variables in a way that is inspired from the LQG-inspired quantization of the underlying symmetry reduced models. More specifically, these modifications consist of polymerisation functions of the connection variables and inverse triad corrections which reflect the quantization in terms of holonomies and fluxes in full LQG. Following \cite{Giesel:2023tsj, Giesel:2024mps} the most general class of effective models considered in \cite{Giesel:2023tsj, Giesel:2024mps} is constructed by encoding these two type of corrections into so-called polymerisation functions $f(K_x, K_\phi, E^x, E^\phi)$ and $h_1(E^x)$, $h_2(E^\phi)$. To construct the underlying spherically symmetric model we restrict to those classes, denoted in \cite{Giesel:2024mps} as class II, which are models that have a compatible LTB condition and in addition the effective geometric part of the Hamiltonian constraint is a conserved quantity. As shown in \cite{Giesel:2023tsj} for models of class II the effective dynamics decouples along the radial direction. Furthermore, since we aim at considering effective spherically symmetric models for which an underlying mimetic Lagrangian exist (models of class III), we focus on the subclass in II that intersects with class III, that are models in II $\cap$ III. To belong to a model of class III requires that no inverse triad corrections are involved as polymerisation function. Such polymerisation functions enter then into the definition of the geometric part of the effective Hamiltonian constraint as \cite{Giesel:2023tsj}
\begin{align}
    C^{(\alpha)}(x) = \frac{E^\phi \sqrt{E^x}}{2G}\left[- (1 + f) \left(\frac{4 K_x K_\phi}{E^\phi} + \frac{K_\phi^2}{E^x}\right) + \frac{1}{E^x} \left(\left(\frac{E^{x\prime}}{2 E^\phi}\right)^2 - 1\right) + 2 \frac{1}{E^\phi} \left(\frac{E^{x\prime}}{2 E^\phi}\right)^\prime\right](x),
\end{align}
where $\alpha$ denotes the polymerisation parameter involved in the function $f$ and we have set $h_1=1$ and $h_2=1$ in the expression in \cite{Giesel:2023tsj} in order that no inverse triad corrections are involved. 
The polymerisation parameter controls
the magnitude of quantum effects and all polymerisation functions are required to recover the correct classical limit 
in the limit $\alpha \to 0$ yielding again the classical Hamiltonian constraint in \eqref{eq:ClassHam}. As shown in \cite{Giesel:2024mps}, the framework is not limited to LQG-inspired polymerisations, but could, in the case of polymerised vacuum solutions, combine models with bounded and unbounded polymerisation functions with specific extended mimetic gravity models within the framework of modified gravity. We refer to polymerised vacuum solutions as the vacuum sector of the effective model, that is $C^{(\alpha)}(x) = 0$, which generally differs from the classical vacuum sector in GR due to quantum gravity corrections.
~\\
~\\
In the case of LQG-inspired models the polymerisation parameter is denoted by $\alpha_\Delta = \gamma \sqrt{\Delta}$ where $\Delta = (4 \pi \ell_\mathrm{P})^2$ is the minimal area gap in terms of the Planck length $\ell_\mathrm{P}$ \cite{Rovelli:1994ge, Ashtekar:1996eg}. Turning now to the gravitational part of the spatial diffeomorphism constraint, a basic assumption of the class of models investigated in \cite{Giesel:2023tsj,Giesel:2023hys,Giesel:2024mps} is that it is kept classical throughout the entire framework in order to gain a better control of the Poisson algebra among the constraints. However, allowing for quantum modifications of the spatial diffeomorphism as well would allow to investigate a broader class of effective models and establishes a closer contact to full LQG. This particular aspect would go beyond the scope of this article, so we reserve this topic for future work.

\section{Asymmetric bounce model}
\label{sec:AsymB}
The formalism in \cite{Giesel:2023tsj,Giesel:2024mps} allows to start with a given cosmological model for each radial coordinate and construct from it the underlying spherically symmetric model, which belongs to a class of spherically symmetric models that decouples along the radial direction. In the main part of this work we will investigate such an example, namely the asymmetric bounce model that was first briefly discussed in section 4.1.2 of \cite{Giesel:2024mps} as an application for the formalism and that will be more in detail investigated here. 
As a first step we adopt the polymerisation function from the effective Thiemann-regularised LQC model as the starting point and discuss the effective spherically symmetric Hamiltonian in its decoupled form. In subsection \ref{sec:IIIED} we will then review the effective dynamics of the model in the marginally bound sector and study some phenomenological implications in \ref{sec:IIIPI} with a particular focus on the formation of shell-crossing singularities. These analytical results are accompanied by numerical investigations of the inhomogeneous dust collapse described by the effective dynamics in subsection \ref{sec:IIINR} together with a final comparison to the well-known symmetric bounce model inspired from standard LQC in \ref{sec:IIISB}.

\subsection{Effective dynamics}\label{sec:IIIED}
The form of the Thiemann-regularised LQC model Hamiltonian was first derived in \cite{Yang:2009fp}, where a Thiemann regularisation of the Euclidean and Lorentzian parts of the Hamilton constraint in full LQG was considered, and then afterwards symmetry reduction to the cosmological sector was performed. A similar result was obtained  in \cite{Dapor:2017rwv} by considering the Euclidean and Lorentzian part of the Hamiltonian constraint in the Algebraic Quantum Gravity (AQG) \cite{Giesel:2006uj} approach to LQG by calculating the semiclassical expectation value of the full AQG Hamiltonian in its graph non-changing regularisation using complexifier coherent states \cite{Thiemann:2000bw, Thiemann:2000ca, Thiemann:2000bx, Thiemann:2002vj} specialised to flat FLRW spacetime and by means of the semiclassical perturbation theory developed in \cite{Giesel:2006um}. 

\begin{figure}[t!]
    \centering
    \includegraphics[width=0.45\linewidth]{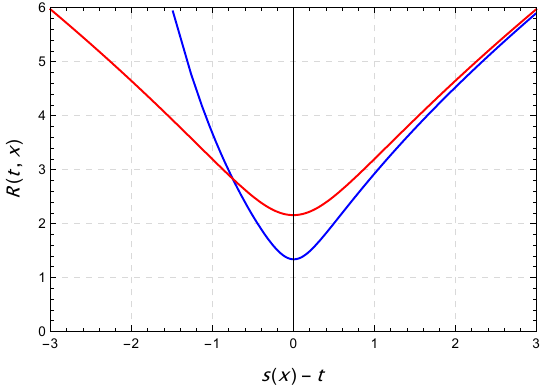}
    \hfill
    \includegraphics[width=0.45\linewidth]{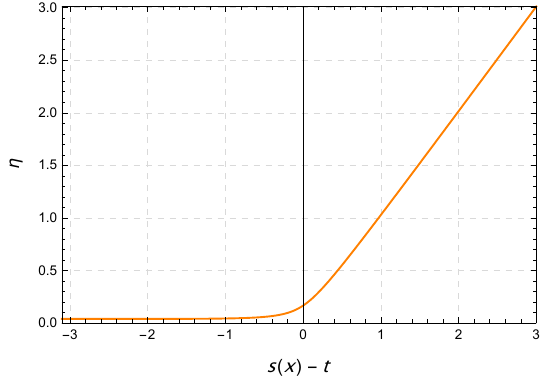}
    \caption{The evolution of the areal radius (left panel) for the mass $M(x) = 5$ in the symmetric (red) and asymmetric (blue) bounce model and the parametric relation between the time coordinate $t$ and the parameter $\eta$ (right panel). }
    \label{fig:RSol}
\end{figure}

Starting to review the construction of the effective spherically symmetric model from \cite{Giesel:2024mps} with an asymmetric bounce we choose the decoupled models along the radial direction to be described by the above discussed asymmetric bounce LQC model. The decoupled effective Hamiltonians $ H^\Delta(x)$ in the marginally bound case can take the form \cite{Giesel:2024mps}
\begin{align}\label{eq:Heff}
    H^\Delta(x) = \frac{v \sin^2(\alpha_\Delta b \gamma)[1 - (1  + \gamma^2) \sin^2(\alpha_\Delta b \gamma)]}{(\alpha_\Delta \gamma)^2},
\end{align}
where we defined the elementary variables $b$ and $v$ as
\begin{align}
    b(t, x) := \frac{K_\phi}{\sqrt{E^x}}(t, x),\hspace{.5cm}v(t, x) := (E^x)^{3/2}(t, x). 
\end{align}
The corresponding effective equations of motion of these variables are then computed to be
\begin{align}
    \Dot{v} &= \{v, H^\Delta(x)\} = \frac{3 v [-\gamma^2 + (1 + \gamma^2) \cos(2 \alpha_\Delta b \gamma)] \sin(2 \alpha_\Delta b \gamma)}{2 \alpha_\Delta \gamma}, \label{eq:EOMv}\\
    \Dot{b} &= \{b, H^\Delta(x)\} = \frac{3 \sin^2(\alpha_\Delta b \gamma) [- 1 + (1 + \gamma^2) \sin^2(\alpha_\Delta b \gamma)]}{2 (\alpha_\Delta \gamma)^2}.
\end{align}
Note that in the above expressions the polymerisation parameter always appears in combination with the Barbero-Immirzi parameter which is required in order to ensure to correct classical limit.
~\\
~\\
The framework developed in \cite{Giesel:2023tsj, Giesel:2024mps} now allows to construct the corresponding gravitational part of the Hamiltonian constraint $C^{(\alpha_\Delta)}(x)$ of the underlying effective spherically symmetric model as well as the corresponding mimetic Lagrangian after the explicit form of the polymerisation function $f$ of this model has been identified. 
While referring the reader for further details on this construction to the work in \cite{Giesel:2023tsj,Giesel:2024mps}, we specifically want to make use of the conservation of $C^{(\alpha_\Delta)}(x)$ in what follows. Using \eqref{eq:Heff} we can eliminate the variable $b$ in the evolution equation \eqref{eq:EOMv} and derive a modified Friedmann equation \cite{Giesel:2024mps}
\begin{align}
\label{eq:ModFRWAsymB}
    \Dot{R} = \pm \frac{x_0}{\sqrt{2} \alpha_\Delta ({\gamma}^2+1) R^2} \sqrt{\frac{\mp x_0 + 4 \alpha_\Delta^2 \left({\gamma}^2+1\right) G M(x)+R^3}{R^3}},
\end{align}
where we used $v(t, x) = R(t, x)^3$ and $x_0 := [R^6-8 \alpha_\Delta^2 {\gamma}^2 \left({\gamma}^2+1\right) G R^3 M(x)]^{1/2}$. The different signs arise due to the fact that the effective decoupled Hamiltonian densities are a quadratic function of $\sin^2$, thus yielding two different branches denoted by $R_\pm$ which are smoothly connected at the bounce where $\Dot{R} = 0$. The implicit solution in its parametric form reads
\begin{align}\label{eq:RSol}
    R(t, x) = \sqrt[3]{\frac{2 G M(x) (4 \alpha_\Delta^2 \gamma^2 + 9 \eta^2)^2}{36 \eta^2 - 16 \alpha_\Delta^2 \gamma^4}},\hspace{.5cm} s(x) - t = \eta - \frac{2}{3} \alpha_\Delta (1 + \gamma^2) \tanh^{-1}\left(\frac{2 \alpha_\Delta \gamma^2}{3 \eta}\right),
\end{align}
where $s(x)$ is a time independent integration constant. In fact, the solution contains both $R_+$ and $R_-$ which are defined on the domains $2 \alpha_\Delta \gamma^2/3 < \eta \le \eta_0$ and $\eta \ge \eta_0$ respectively. The bounce is located at $\eta_0 := 2/3 \gamma \alpha_\Delta (1 + 2 \gamma^2)^{1/2}$ where the areal radius attains its minimal value $R(\eta_0) = 2 [\alpha_\Delta^2 \gamma^2 (\gamma^2 + 1) G M(x)]^{1/3}$. We choose $R_-$ to describe the pre-bounce phase where the dynamics of the collapse starts at some initial time since the above parametric relation implies $t \to -\infty$ in the limit $\eta \to \infty$. Accordingly, $R_+$ is then associated with the post-bounce dynamics following from the fact that $t \to \infty$ as the parameter $\eta$ reaches the lower bound of the respective interval. This ensures that that we obtain the classical solution asymptotically in the pre-bounce phase and asymptotically de-Sitter in the post-bounce phase. To this end, we want to remark that from \eqref{eq:RSol} the classical LTB solution is consistently recovered in the limit $\alpha_\Delta \to 0$ as $\eta \to s(x) - t$. 
~\\
~\\
The evolution of the areal radius for the fixed mass $M(x) = 5$ as well as the parametric relation are depicted in Fig.~\ref{fig:RSol}. Note that for illustrative purposes we set the polymerisation parameter $\alpha_\Delta = 1$ in Planckian units while the Barbero-Immirzi parameter takes the value $\gamma = 0.2375$ suggested from results in the context of black hole entropy within symmetry reduced models of LQG \cite{Ashtekar:1997yu}, which corresponds here to the illustrative choice $\ell_p=\frac{1}{\sqrt{0.2375}\sqrt{4\pi}}$. As can be seen in the left panel of Fig.~\ref{fig:RSol}, the shape of the blue graph relative to the evolution of the areal radius from the symmetric bounce model outlined in red is asymmetric around the vertical axis and thus reflects the impact of the additional Lorentzian term in the effective Hamiltonian \cite{Dapor:2017rwv, Li:2018opr}. These differences become further apparent at the level of equations of motion when considering the large volume limit as in the cosmological case \cite{Assanioussi:2018hee}. For the branch $R_+$ one finds that the modified Friedmann equation can be reduced to $[\Dot{v}/(3v)]^2 = \Lambda / 3$ where $\Lambda := 3 / [\alpha_\Delta^2 (\gamma^2 + 1)^2]$ is interpreted as an emergent (Planckian order) cosmological constant. As a consequence, the post-bounce phase of this model is described by a (effective) de Sitter universe for large $t$, while $R_-$ on the other hand corresponds the standard flat FLRW spacetime as in the case of the symmetric bouncing scenario.

\newpage

\subsection{Phenomenological implications}\label{sec:IIIPI}

The qualitative differences in the effective dynamics that were discovered first in the cosmological context for instance in \cite{Dapor:2017rwv, Li:2018opr, Assanioussi:2018hee} also hold in the spherically symmetric case as by construction for the class of models considered here each decoupled shell is described by an effective cosmological model. In this subsection, we focus on the phenomenological implications of the effective asymmetric bounce model,  with emphasis  on the existence and/or resolution of singularities. This concerns the central singularity in connection with the formation of black holes in the polymerised vacuum sector, the formation of (anti-)enclosed surfaces, and, in particular, the formation of shell-crossing singularities. The latter will be further analysed in future work \cite{tba}. 

\subsubsection{Strong and weak singularities}

We begin with the bouncing behaviour of the effective dynamics which in general is well-known to remove the central singularity in effective models due to the quantum geometric modifications expected to be dominant at Planckian scales. First indications closely connected to the fact that the areal radius of each shell has a non-trivial minimum value at the bounce as can be seen from the modified Friedmann equation in \eqref{eq:ModFRWAsymB} where the explicit minimal value is discussed below \eqref{eq:RSol}.
Inserting the solution from \eqref{eq:RSol} and evaluating the expression at the bounce $\eta = \eta_0$ leads to the critical energy density
\begin{align}
    \rho_\mathrm{crit} = \frac{3}{32 \pi G \alpha_\Delta^2 \gamma^2 (1+\gamma^2)},
\end{align}
which exactly coincides with the one obtained in the cosmological context \cite{Assanioussi:2019iye}. In particular, it is independent of the mass function $M(x)$ as a consequence of the so-called $\overline{\mu}$ quantization scheme from LQC \cite{Ashtekar:2006wn} apparent in the polymerisation function of this model. Another way of showing that strong singularities are absent is by calculating the curvature scalars using the explicit expression for the areal radius. The Ricci scalar and the Kretschmann scalar for a generic mass function $M(x)$ read
\begin{align}\label{eq:curv}
    \mathcal{R} = \frac{\mathcal{A}}{(9 \eta^2 + 4 \alpha_\Delta^2 \gamma^2)^4 \mathcal{S}},\hspace{.5cm}\mathcal{K} = \frac{\mathcal{B}}{(9 \eta^2 + 4 \alpha_\Delta^2 \gamma^2)^8 \mathcal{S}^2},
\end{align}
where the explicit form of $\mathcal{A}$ and $\mathcal{B}$ are provided in appendix \ref{A:Curv} and the functional $\mathcal{S}$ in the denominator is given by
\begin{align}\label{eq:S}
    \mathcal{S} := -18 \eta M(x)[4 \alpha_\Delta^2 (2 \gamma^4 + \gamma^2) - 9 \eta^2]s'(x) + (4\alpha_\Delta^2 \gamma^2 + 9 \eta^2)^2 M'(x).
\end{align}
Its explicit value at the bounce takes the form $\mathcal{S}(\eta_0) = 64 \alpha_\Delta^4 \gamma^4 (\gamma^2 + 1)^2 M'(x)$ which is clearly non-vanishing within the dust interior where $M^\prime(x) \neq 0$. Consequently, the curvature remains finite at the bounce for $\alpha_\Delta \neq 0$, confirming that the central singularity is resolved due to quantum corrections, likewise to the case of the asymmetric bounce model in cosmology. In the polymerised vacuum region for which $M^\prime(x) = 0$, however, the curvature invariants become ill-defined at $\eta = \eta_0$ and it can be shown that they diverge with respect to $\eta$ as $1/\eta$ and $1/\eta^2$, respectively.
Such a power law behaviour is different than for the central singularity in the classical case and we identify it with the presence of shell-crossing singularities where $R'=0$. This is in contrast to the symmetric bounce model where no weak singularities are present in the polymerised vacuum region. A possible explanation for this occurrence that will be explored in future works is the non-trivial additional contribution due to quantum gravity corrections that could in principle be moved to the right-hand side of the Einstein's equations in the case of the polymerised vacuum  solution which is sourced by higher order derivative couplings of the mimetic scalar field, see the discussion in section 3.1.3 of \cite{Giesel:2024mps}.
A more detailed investigation of the existence of shell-crossing singularities not only for the asymmetric bounce model, but also for models with unbounded polymerisation functions, which have been shown to be related to modified black hole models such as that of Bardeen and Hayward, is presented in \cite{tba}. There, the question of the existence of shell-crossing singularities was analysed following \cite{Fazzini:2023ova}, where only the symmetric bounce model was considered, and the results in \cite{tba} show that the question of the existence of shell-crossing singularities can be further reduced to a simplified inequality condition for the quantities $M(x)$, $s(x)$ and their derivatives, depending on whether the initial profile is increasing or decreasing.
~\\
~\\
Finally we turn to the asymptotic behaviour of the curvature invariants and find that in the limit $\eta \to 2 \alpha_\Delta \gamma^2 / 3$ or equivalently $t \to \infty$ by the parametric relation in \eqref{eq:RSol} we have
\begin{align}\label{eq:limRic}
    \lim_{t \to \infty} \mathcal{R} = \frac{12}{\alpha_\Delta^2 (\gamma^2 + 1)^2} = 4 \Lambda,
\end{align}
whereas for $\eta \to \infty$ or $t \to -\infty$ the curvature converges to zero. The above value exactly matches the curvature scalar of the de Sitter spacetime in four dimensions and thus supports the above observation concerning the late time evolution of the areal radius. Furthermore, we see from \eqref{eq:limRic} that the curvature remains on the order of Planckian scales throughout the entire dynamics after the bounce, which is a characteristic feature of this effective model.

\subsubsection{Formation of (anti-)trapped surfaces}
\begin{figure}[t!]
    \centering
    \includegraphics[width=0.45\linewidth]{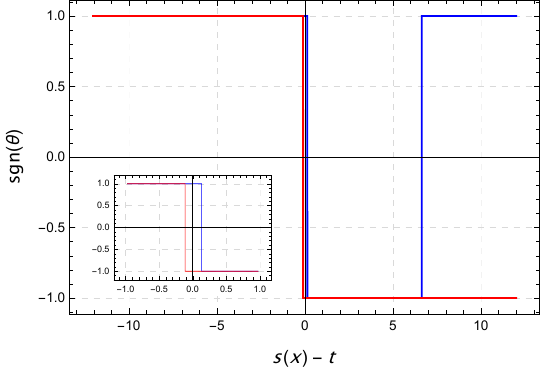}
    \hfill
    \includegraphics[width=0.45\linewidth]{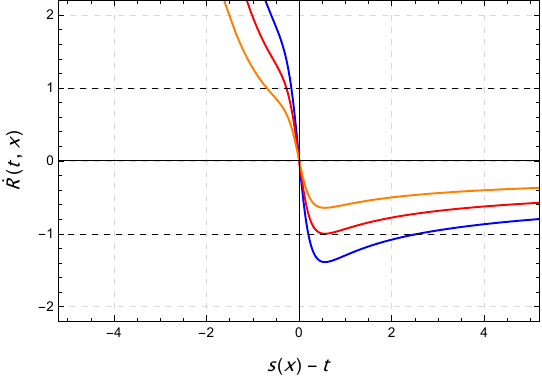}
    \caption{The signs of the expansion parameters $\theta_+$ (blue) and $\theta_-$ (red) plotted over $s(x) - t$ for the mass $M(x) = 5$ (left panel) and the time derivative of the solution $R(t, x)$ plotted over $s(x) - t$ for the masses $M(x) = 0.2, 0.747326, 2$ (orange, red, blue) indicating the existence of a critical mass for the formation of trapped regions (right panel). The inner diagram in the left panel shows an enlarged view of a small region around $s(x) - t = 0$. }
    \label{fig:trapped}
\end{figure}

Next we investigate the formation of trapped and anti-trapped regions during the gravitational collapse and expansion of the dust sphere. For a family of incoming and outgoing null congruences that meet orthogonally at a surface with constant surface radius, we can define the corresponding future-directed null vector fields as follows:
\begin{align}
    X_+ = \frac{1}{\sqrt{2}}\left(\partial_t + \frac{1}{R^\prime} \partial_x\right),\hspace{.5cm}X_- = \frac{1}{\sqrt{2}}\left(\partial_t - \frac{1}{R^\prime} \partial_x\right).
\end{align}
Then a trapped surface is said to form if the radius of the sphere shrinks along the radial null geodesics \cite{Hayward:1994bu}. This can be quantified by introducing the respective expansion parameters which for the model under consideration read
\begin{align}\label{eq:thetapm}
    \theta_\pm = \frac{\sqrt{2}}{R^\prime} (\partial_t R \pm 1).
\end{align}
Accordingly, for trapped (anti-trapped) spheres we have that $\theta_+ \theta_- > 0$ and $\theta_\pm < 0$ ($\theta_+ \theta_- > 0$ and $\theta_\pm > 0$) whereas for untrapped spheres $\theta_+ \theta_- < 0$ holds. We thus obtain a condition $\theta_+ \theta_- = 0$ or equivalently $\Dot{R} = \pm 1$ allowing us to identify the coordinate location of the apparent horizons.
~\\
~\\
The formation of trapped and anti-trapped regions during the evolution of the dust sphere can be visualised by calculating the expansion parameters explicitly using the solution in \eqref{eq:RSol}. In Fig. \ref{fig:trapped} the corresponding signs of the parameters are plotted over $s(x) - t$ where for illustrative purposes we again set $\alpha_\Delta = 1$, $\gamma = 0.2375$ and $M(x) = 5$. From the arrangement of the blue and red curves, which correspond to the signs of $\theta_+$ and $\theta_-$, we find that considering a fixed dust shell it is trapped at a certain point until it crosses the inner horizon $\theta_+ = 0$ shortly before the bounce where $s(x) - t \approx 0$ applies for the given values. Just as in the symmetric bouncing scenario for which the details were worked out for instance in \cite{Giesel:2022rxi}, this indicates that a black hole is forming while the dust sphere is collapsing. After the shell has crossed the untrapped region around the bounce shown by the inner diagram in the left panel of Fig. \ref{fig:trapped}, it enters the inner horizon corresponding to $\theta_- = 0$ and remains anti-trapped during the entire post-bounce evolution from there on. This is a distinct feature of the asymmetric bounce model and thus leads to the interpretation of a cosmological (de-Sitter) horizon. Interestingly, a similar conclusion can be found in \cite{Ou:2025bbv} in the special case of the effective Oppenheimer-Snyder collapse for which the contribution of the Lorentzian part of the constraint is taken into account as well. Our results show that a certain critical mass must be exceeded in order for a black hole horizon to form as illustrated in the right panel of Fig. \ref{fig:trapped} for the masses $M(x) = 0.2, 0.747326, 2$, a similar property that was previously obtained for effective LTB collapse models based on the symmetric bounce model \cite{Kelly:2020uwj,Giesel:2021dug,Giesel:2022rxi,Lewandowski:2022zce,Giesel:2023hys}.
In the numerical analysis performed we have that for $M > M_\mathrm{c} = 0.747326$ the function $\Dot{R} + 1$ contains two roots corresponding to the inner and outer horizon in the pre-bounce phase. On the other hand, the condition $\Dot{R} - 1 = 0$ is satisfied only for one value of the time coordinate and irrespective of the value of the dust mass. 

\subsection{Numerical results of the inhomogeneous dust collapse}\label{sec:IIINR}

\begin{figure}[t!]
    \centering
    \includegraphics[width=0.45\linewidth]{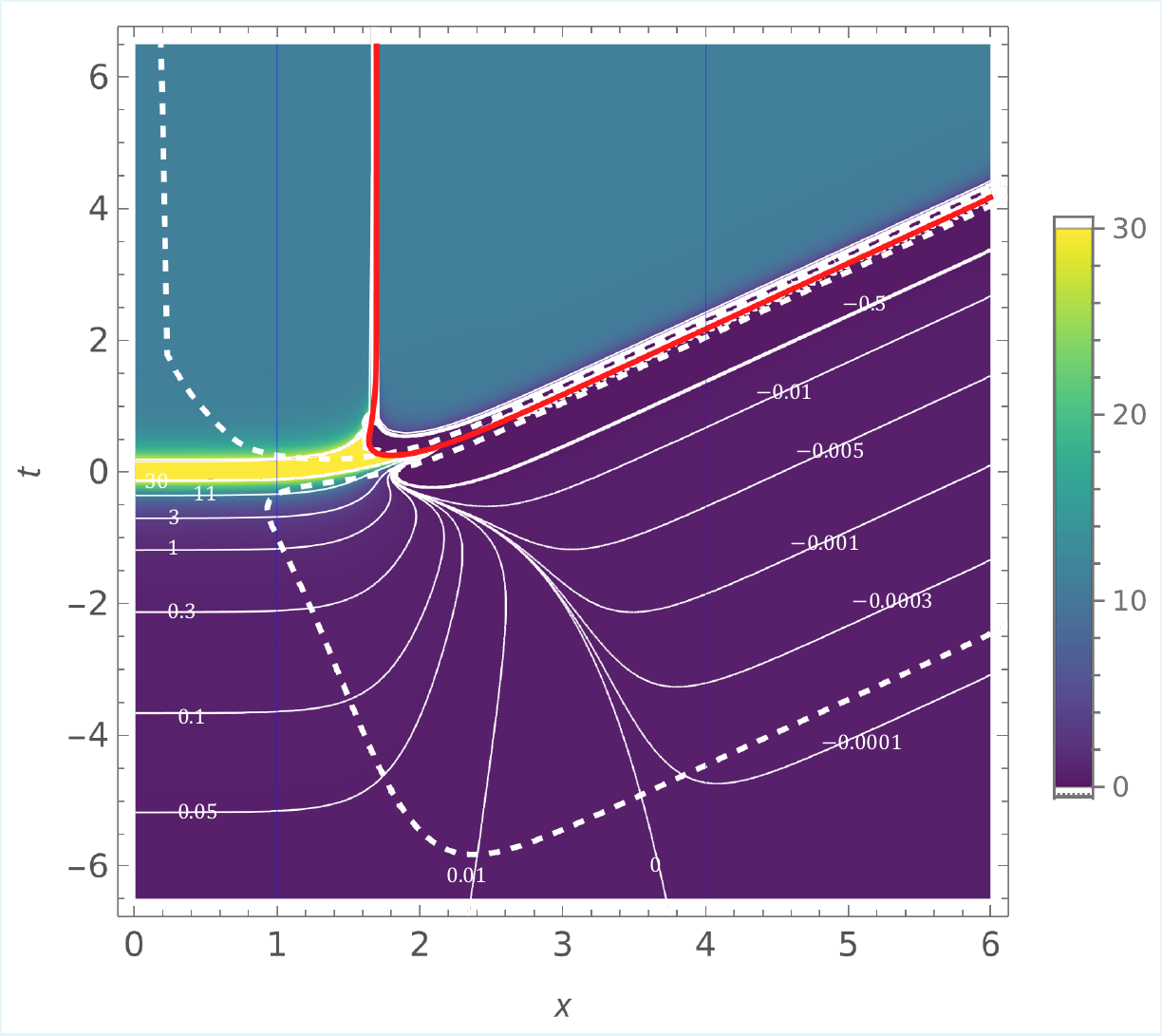}
    \hfill
    \includegraphics[width=0.45\linewidth]{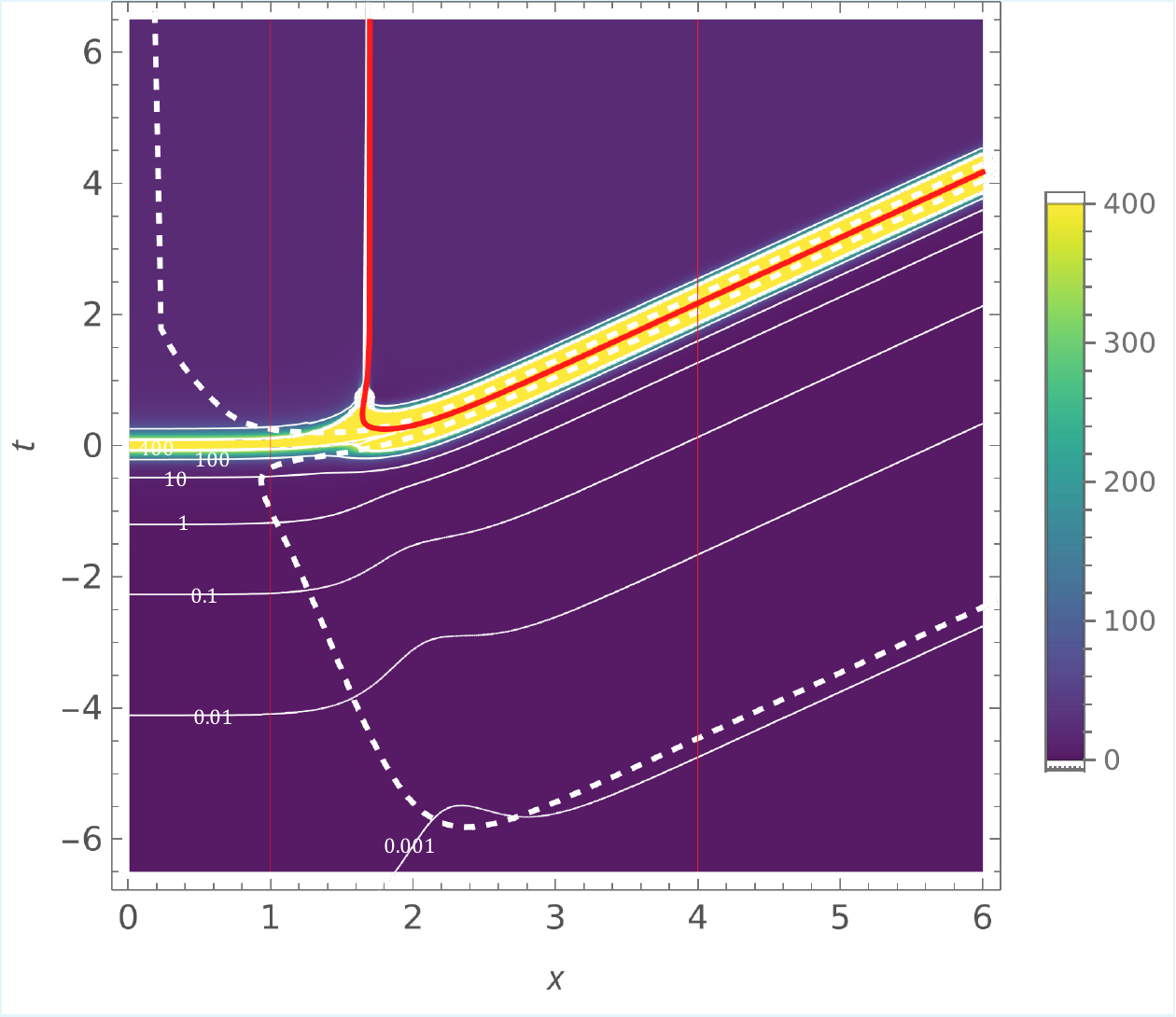}
    \caption{The Ricci scalar (left figure) and the Kretschmann scalar (right figure) shown over the $t - x$ plane. The red line indicates the spacetime location of the shell-crossing singularity and the white dashed lines mark the apparent horizons. The vertical lines at $x = 1$ and $x = 4$ display the position where the cross-sections for the diagrams provided in Fig. \ref{fig:curvature1D} were taken.}
    \label{fig:curvature2D}
\end{figure}

\begin{figure}[t!]
    \centering
    \includegraphics[width=0.45\linewidth]{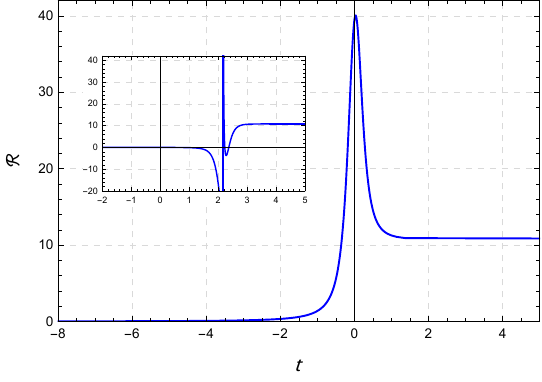}
    \hfill
    \includegraphics[width=0.45\linewidth]{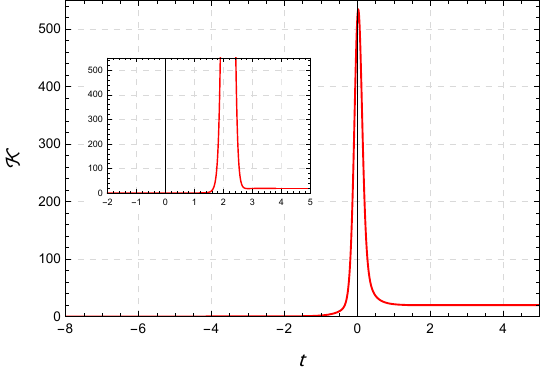} \\
    \vspace{.3cm}
    \includegraphics[width=0.5\linewidth]{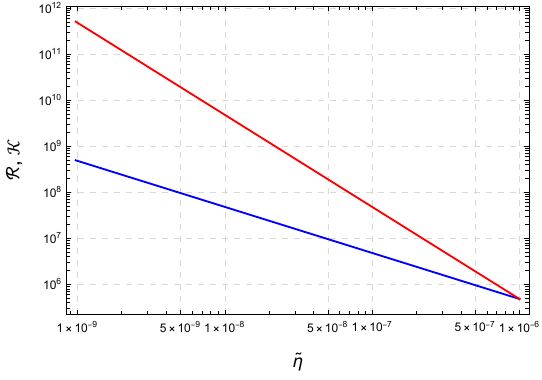}
    \caption{The top left and right panels show the time evolution of the Ricci and Kretschmann scalar for fixed radial coordinates $x = 1$ (outer diagram) and $x = 4$ (inner diagram). The bottom diagram displays the curvature scalars at $x = 4$ plotted near the singularity on a double logarithmic scale where $\widetilde{\eta} := \eta - \eta_0$.}
    \label{fig:curvature1D}
\end{figure}

In this section we perform a numerical analysis of the inhomogeneous dust collapse and investigate its dynamical properties in the marginally bound case based on the theoretical results established in the previous sections. For this purpose, we will choose physically reasonable profiles for the integration constant $s(x)$ and mass function $M(x)$ involved in the dynamical equations which were treated as yet unspecified time-independent functions. These profiles are then substituted into the expressions for the physical quantities such as the curvature scalars and expansion parameters which are subsequently evaluated numerically. Similar to the results presented in the previous sections, we will thereby fix the polymerisation parameter $\alpha_\Delta = 1$ and the Immirzi parameter $\gamma = 0.2375$. 
\begin{figure}[t!]
    \centering
    \includegraphics[width=0.45\linewidth]{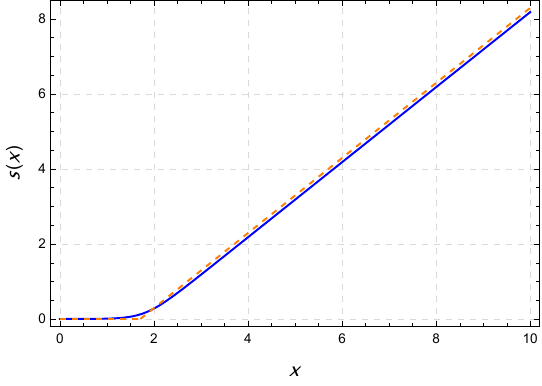}
    \hfill
    \includegraphics[width=0.45\linewidth]{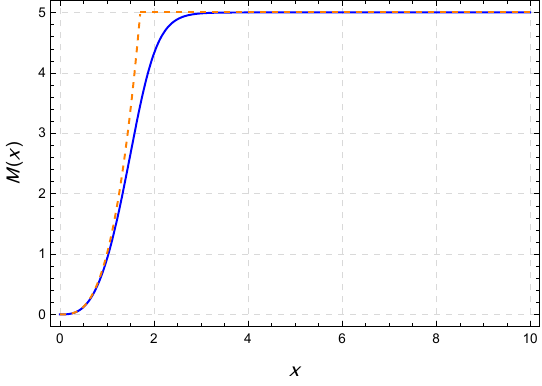}
    \caption{Inhomogeneous profiles (blue) for the integration constant $s(x)$ (left panel) and the mass function $M(x)$ with a maximum value of $M = 5$ in the polymerised vacuum region (right panel) based on the homogeneous reduction to the Oppenheimer-Snyder model (orange dashed).}
    \label{fig:profiles}
\end{figure}
~\\
~\\
We begin with a specific choice for the time independent function $s(x)$ which arises when solving the equation of motion \eqref{eq:RSol} for the areal radius and can be understood as an integration constant in each of the decoupled equations of motion along the radial coordinate. As regards the effective LTB line element, the integration constants encode the residual gauge freedom of selecting  the radial coordinate, since different choices all lead to the same line element. In particular, we can set $s(x) = x$ when we reach the outer polymerised vacuum region of the dust sphere, since in this case the Schwarzschild metric in LTB coordinates is restored upon rescaling. Furthermore, we assume that a homogeneous reduction to FLRW is possible within the dust sphere, allowing us to set $s(x)$ as a constant function. In order for the bounce to appear at $t \approx 0$ for the chosen values of $\alpha_\Delta$ and $\gamma$, it is convenient work with $s(x) = 0$ in this region. For the mass distribution $M(x)$, we can model the interior of the dust using $M(x) \sim x^3$, based on the aforementioned reduction to FLRW cosmology, and consider, for example, $M = 5$ as the maximum value after the transition to the Schwarzschild-like polymerised vacuum region. This reduction ansatz leads to the effective Oppenheimer-Snyder model \cite{Giesel:2022rxi} as can been seen from the orange dashed graphs depicted in the two diagrams in Fig. \ref{fig:profiles} and can be generalised to the inhomogeneous case indicated by the blue graph by modelling a suitable function that resembles both regions and is differentiable across the junction.
~\\
~\\
Next, we numerically evaluate the curvature scalars defined in \eqref{eq:curv} as functions of the temporal and radial coordinate $(t, x)$. The resulting plots over the $t - x$ plane can be found in Fig. \ref{fig:curvature2D} where the contour lines correspond to the current numerical value of the curvature scalars. In addition, we show the cross-sectional plots of the curvature scalars at different radial positions $x = 1, 4$ marked as vertical lines in Fig. \ref{fig:curvature2D} and given in the top left and right panel in Fig. \ref{fig:curvature1D} in order to track the dynamical evolution in the interior and exterior region individually.
~\\
Starting from radial coordinates corresponding to the interior of the dust, we can observe that the curvature scalars reach a precisely defined maximum value at the bounce $t \approx 0$ and then drop to a constant value given by the scalar curvature of de Sitter spacetime. For larger values of the radial coordinate, however, where we enter the Schwarzschild-like polymerised vacuum region, a shell-crossing singularity appears at the bounce $\eta  = \eta_0$, indicated by the continuous red lines in Fig. \ref{fig:curvature2D}, which is consistent with the theoretical results from the previous section. Around the coordinates where the boundary of the dust sphere is located shell-crossing singularities form after the bounce has appeared for $\eta < \eta_0$ reaching to future infinity almost a fixed radial coordinate which is a feature also present in the symmetric bounce model, see also section \ref{sec:IIISB} for a comparison of these two models. This follows from the fact that divergences of the curvature scalar can also occur in a more complex manner for $M'(x) \neq 0$, whereby the function $\mathcal{S}$ in its complete form is a fourth-order polynomial in the parameter $\eta$. More concretely, from the above choice for $M(x)$ leading to a decreasing energy density profile we have that due to the inhomogeneity of this particular region outer collapsing matter shells overlap with the inner ones that have already undergone a bounce as a consequence of their relative velocities. 
~\\
~\\
The dynamical behaviour of the curvature scalars qualitatively described in the previous paragraph for both the interior and exterior region can also be inferred from the cross-sectional plots in Fig. \ref{fig:curvature1D} where the inner and outer plots correspond to the radial coordinates $x = 4$ and $x = 1$ respectively. The divergence of the curvature scalars visible in the case of $x = 4$ can be further analysed by plotting one of the two diverging branches on a double logarithmic scale as shown in the lower panel of Fig. \ref{fig:curvature1D} where we defined $\widetilde{\eta} := \eta - \eta_0$ and rescaled the Kretschmann scalar given by the red curve for illustrative purposes. From the linear shape of the graphs it follows that the curvature scalars exhibit the power-law behaviour $1/\eta$ and $1/\eta^2$ up to tiny numerical errors. This holds also in the case where at a given instant of time after the bounce the curvature is varied instead over the radial coordinate $x$. We thus find a numerical evidence for this particular effective model that the development of shell-crossing singularities for physically reasonable profiles as displayed in Fig. \ref{fig:profiles} is a typical feature. For a more detailed discussion on different choices of initial profiles the reader is referred to \cite{tba}.
~\\
~\\
Finally, we turn to the formation of trapped surfaces and apply the results obtained from the previous section to this numerical example. The coordinate locations for which the expansion parameters vanish are indicated by the white dashed lines in the curvature plots in Fig. \ref{fig:curvature2D}. When a certain threshold value of the mass function $M(x)$ is exceeded, i.e. as the values for the radial coordinate increase, there is an inner and an outer horizon in the pre-bounce phase, which enclose a region of spacetime in which the dust shells are trapped. Compared to the symmetric bounce model, the novelty of the asymmetric bounce model is particularly evident in this respect in the post-bounce phase, in which the radial region of the inner horizon is not limited by a critical mass as in the symmetric bounce model. This means that each individual dust shell eventually ends up in an anti-trapped region that remains in place throughout the entire post-bounce dynamics.

\subsection{Comparison to the symmetric bouncing scenario}\label{sec:IIISB}

In this section, we conclude the previous numerical analysis with a concrete comparison to the symmetric bounce model, which has been used in the literature for both homogeneous dust profiles in connection with the Oppenheimer-Snyder collapse \cite{Lewandowski:2022zce, Giesel:2022rxi, Fazzini:2023scu} and for inhomogeneous configurations \cite{Giesel:2023hys, Fazzini:2023ova}. In the following, we restrict ourselves specifically to the latter case for the marginally bound model and follow the same steps as in Section \ref{sec:IIINR}, applying the dust profile  and the integration constant defined above to this specific model.
~\\
~\\
The solution for the effective marginally bound model explicitly reads \cite{Giesel:2023hys}
\begin{align}\label{eq:RSolSym}
    R(t, x) = \left(2 G M(x) \left(\frac{9}{4} (s(x) - t)^2 + \alpha_\Delta^2\right)\right)^{1/3},
\end{align}
where $\alpha_\Delta$ is the same polymerisation parameter as above, related to the minimum surface distance $\Delta$. In the limiting case $\alpha_\Delta \to 0$ the classical LTB solution for the marginally bound case is rediscovered. As already shown in Fig. \ref{fig:RSol} for the same choice of $\alpha_\Delta$ and $M(x)$ as in the asymmetric bounce model, the solution describes a bounce with a symmetric shape around $s(x) - t = 0$, where the minimum radius is given by $R_\mathrm{min} = (2 \alpha_\Delta^2 G M(x))^{1/3}$. By defining $z := s(x) - t$, the Ricci and Kretschmann scalars can be calculated directly as \cite{Giesel:2024mps}
\begin{align}\label{eq:curvSymm}
    \mathcal{R} = \frac{\mathcal{C}}{(9 z^2 + 4 \alpha_\Delta^2)^2 \mathcal{T}},\hspace{.5cm}\mathcal{K} = \frac{\mathcal{D}}{(9 z^2 + 4 \alpha_\Delta^2)^4 \mathcal{T}^2},
\end{align}
where $\mathcal{C}$, $\mathcal{D}$ are given in the appendix \ref{A:Curv} and $\mathcal{T} := M'(x) (9 z^2 + 4 \alpha_\Delta^2) + 18 M(x) s'(x) z $. While the strong singularity at the bounce $z = 0$ is resolved due to the presence of the polymerisation parameter, shell-crossing singularities may still occur if $\mathcal{T}$ vanishes for real $z$. However, it should be noted  that in the polymerised vacuum case $M'(x) = 0$ the curvature scalars remain finite since both the numerator and denominator become proportional to the same powers of $z$ which ultimately cancel out \cite{Giesel:2024mps}.
~\\
~\\
\begin{figure}[t!]
    \centering
    \includegraphics[width=0.45\linewidth]{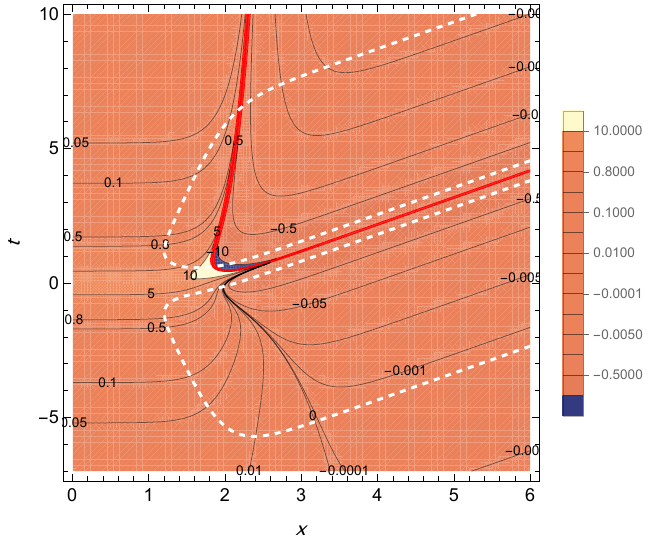}
    \hfill
    \includegraphics[width=0.45\linewidth]{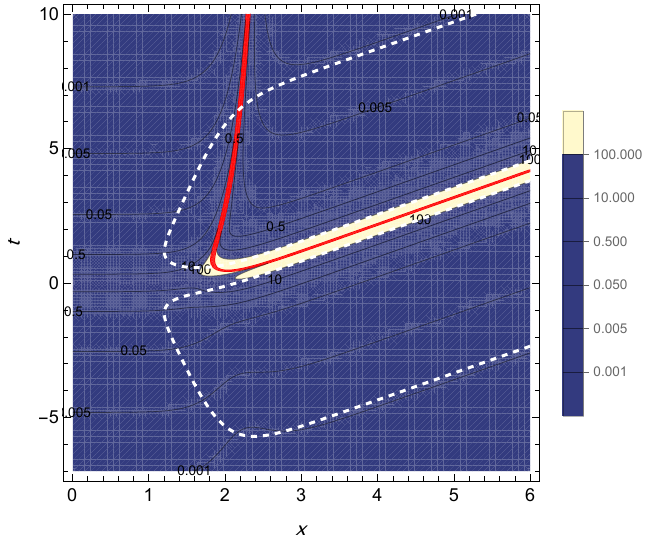}
    \caption{The Ricci scalar (left figure) and the Kretschmann scalar (right figure) in the symmetric bounce model plotted over the $t - x$ plane. The red line shows the spacetime location of the shell-crossing singularity and the white dashed lines the apparent horizons.}
    \label{fig:curvature2Dsymm}
\end{figure}
For the qualitative comparison to the asymmetric bounce model we now numerically compute the curvature scalars as well as the expansion parameters using the profiles for $s(x)$ and $M(x)$ constructed in section \ref{sec:IIINR}. The resulting two-dimensional plots over the $t - x$ plane can be found in Fig. \ref{fig:curvature2Dsymm}. First we notice that the dynamical evolution of the curvature in the pre-bounce phase is identical to the asymmetric case shown in Fig. \ref{fig:curvature2D} up to different scales close to the bounce which follows directly from the same asymptotic limit described by the FLRW spacetime in both cases. However, when passing to the post-bounce regime we realise that the curvature scalars decrease symmetrically around the bounce to zero instead of reaching a non-vanishing constant value which is simply a consequence of the axial symmetry of the areal radius given in \eqref{eq:RSolSym}.
~\\
~\\
Turning now to the formation of shell-crossing singularities, the plots in Fig. \ref{fig:curvature2Dsymm} show that similar to the asymmetric case in Fig. \ref{fig:curvature2D} the red lines marking their locations in the $t-x$ plane extend to the polymerised vacuum region which seems to be in contradiction with the explicit form of the curvature scalars in \eqref{eq:curvSymm}. The reason for this particular behaviour is that the constructed mass function displayed in Fig. \ref{fig:profiles} by the blue curve is not truly a constant function at large values of $x$ where the polymerised vacuum region is assumed. Instead, $M(x)$ slowly converges to the chosen maximum value $M = 5$ with a non-vanishing derivative $M'(x)$ causing the curvature scalars still to diverge approximately at the bounce $s(x) - t$. However, for the qualitative discussion this poses no further obstruction. Another important difference concerns the formation of anti-trapped surfaces in the post-bounce region. Compared to the plots in Fig. \ref{fig:curvature2D} where the anti-trapped region is ubiquitous and not constrained by a critical mass as demonstrated in Fig. \ref{fig:trapped}, there exists an outer horizon enclosing an anti-trapped white hole region after a finite amount of time. Since the deep interior of the dust sphere is untrapped both in the pre- and post-bounce phase throughout the entire evolution, anti-trapped and trapped surfaces only form once a critical mass is passed which in contrast to the asymmetric case can be analytically obtained as $M_\mathrm{c} = 8 \alpha_\Delta / (3 G \sqrt{3})$ \cite{Kelly:2020uwj,Lewandowski:2022zce,Giesel:2021dug,Giesel:2023hys}. 

\section{Conclusion}
In this work, we have investigated the effective LTB model with an  asymmetric bounce for inhomogeneous dust profiles, focusing in particular on its phenomenological properties. This extends previous work on the effective LTB model with symmetric bounce \cite{Bojowald:2008ja,Bojowald:2009ih,Munch:2020czs, Husain:2021ojz,Giesel:2021dug,Giesel:2022rxi,Lewandowski:2022zce,Fazzini:2023ova,Cipriani:2024nhx} as well as more recent work on the asymmetric bounce model for homogeneous profiles \cite{Ou:2025bbv}. As already evident from the analytical results in \cite{Giesel:2023hys}, the numerical results for a specific inhomogeneous dust profile confirm the properties that the central singularity at the bounce in this model is resolved for profiles whose mass function satisfies $M'(x)\not=0$, i.e. inside the dust. For the outer polymerised vacuum region, where $M'(x)=0$ at the bounce, there is a singularity that we have identified as a shell-crossing singularity. The reason for this is that the power-law behaviour of the curvature scalar values is different from what is expected for a strong singularity, which we have confirmed through both analytical and numerical results. This is a difference from the effective LTB model with symmetric bounce, in which there are no shell-crossing singularities in the polymerised vacuum region.
~\\
~\\
Our analysis of the formation of trapped and anti-trapped surfaces during the evolution of the dust sphere showed that above a critical mass value in the collapsing pre-bounce phase, there is an inner and outer horizon, including a region of spacetime in which the dust shells are trapped, similar to the effective LTB model with symmetric bounce. While this critical mass can be calculated analytically in the model with a symmetric bounce, we were only able to determine it numerically here. In contrast, the inner horizon in the post-bounce phase is not subject to any critical mass restrictions, and the anti-trapped region is present throughout the post-bounce evolution, which is a significant difference from the effective LTB model with a symmetric bounce.
~\\
~\\
While in this work we have focused on investigations of shell-crossing singularities of the effective LTB model with a symmetric bounce for a specific choice of dust profile and integration constants, in \cite{tba} we will extend this analysis to derive more general conditions for the existence of shell-crossing singularities  for this model. In addition such analysis will also include models with unbounded polymerisation functions, which can be related to the regular black hole models of Bardeen \cite{Bardeen68} and Hayward \cite{Hayward:1994bu} for particular choices of these functions. Another interesting future line of research will be to better understand the extension of spacetime beyond shell-crossing singularities in terms of weak solutions or thin-shell models, which have so far only been applied to the effective LTB model with a symmetric bounce. Furthermore, recent results in \cite{Giesel:2025kdl} allow for a more detailed investigation of the non-marginally bound effective LTB model and the question of the  absence or presence of shell-crossing singularities, which we intend to do in future work. Furthermore, a more detailed investigation of the quantum models underlying these effective LTB models and a comparison with quantum models formulated in Schrödinger quantisation \cite{Kiefer:2019csi} would be an interesting topic for future research.
\label{sec:Concl}

\begin{acknowledgments}
E.R. thanks the Villigst foundation for financial support. K.G. and E.R. would like to thank Francesco Fazzini for valuable discussions about the results of this work.
\end{acknowledgments}
\newpage

\appendix

\section{Curvature invariants}\label{A:Curv}
In the following we list the explicit expressions entering the curvature scalars given in \eqref{eq:curv} and \eqref{eq:curvSymm} for the asymmetric and symmetric bounce model respectively.
~\\
~\\
{\underline{Asymmetric bounce:}}
\begin{align}
        \mathcal{A} &= 36\Bigg(
        M'(x)(4\gamma^2\alpha_{\Delta}^2 + 9\eta^2)^2
        \Big(540(\gamma^2 + 2)\gamma^2\eta^4\alpha_{\Delta}^2 + 64(2\gamma^2 + 1)\gamma^8\alpha_{\Delta}^6
\nonumber\\
&
- 48(2\gamma^4 + 8\gamma^2 - 1)\gamma^4\eta^2\alpha_{\Delta}^4 + 243\eta^6
\Big) + 48\gamma^2\eta\alpha_{\Delta}^2 M(x) s'(x) \nonumber\\
&\times
\Big(
324(2\gamma^4 + 9\gamma^2 + 4)\gamma^2\eta^4\alpha_{\Delta}^2 + 64(7\gamma^4 + 4\gamma^2 - 2)\gamma^8\alpha_{\Delta}^6
\nonumber\\
&
- 144(\gamma^4 + 18\gamma^2 + 14)\gamma^6\eta^2\alpha_{\Delta}^4 - 729(3\gamma^2 + 4)\eta^6
\Big)
\Bigg),\\
        \mathcal{B} &= 432 \Bigg(
384\eta M(x) M'(x) s'(x)
\left(4\gamma^3\alpha_{\Delta}^3 + 9\gamma\eta^2\alpha_{\Delta}\right)^2
\nonumber\\
&\times
\Big(
1024(22\gamma^6 + 27\gamma^4 + 6\gamma^2 - 1)\gamma^{16}\alpha_{\Delta}^{12}
\nonumber\\
&
+ 243\eta^6\Big[
-162\gamma^2(64\gamma^4 + 143\gamma^2 + 72)\eta^4\alpha_{\Delta}^2
\nonumber\\
&
+ 36\gamma^4(291\gamma^6 + 890\gamma^4 + 733\gamma^2 + 144)\eta^2\alpha_{\Delta}^4
\nonumber\\
&
- 16\gamma^6(261\gamma^8 + 1211\gamma^6 + 1479\gamma^4 + 525\gamma^2 + 16)\alpha_{\Delta}^6 + 729(3\gamma^2 + 4)\eta^6
\Big]
\nonumber\\
&
- 768(172\gamma^8 + 515\gamma^6 + 381\gamma^4 + 65\gamma^2 + 1)\gamma^{12}\eta^2\alpha_{\Delta}^{10}
\nonumber\\
&
+ 1728(82\gamma^8 + 743\gamma^6 + 1299\gamma^4 + 641\gamma^2 + 43)\gamma^{10}\eta^4\alpha_{\Delta}^8
\Big)
\nonumber\\
&
+ M'(x)^2 \Big(
4096(2\gamma^2 + 1)^2\gamma^{16}\alpha_{\Delta}^{12} + 81\eta^6\Big[
-648\gamma^2(29\gamma^2 + 22)\eta^4\alpha_{\Delta}^2
\nonumber\\
&
+ 432\gamma^4(95\gamma^4 + 164\gamma^2 + 74)\eta^2\alpha_{\Delta}^4
\nonumber\\
&
- 256\gamma^6(97\gamma^6 + 225\gamma^4 + 144\gamma^2 + 11)\alpha_{\Delta}^6 + 3645\eta^6
\Big]
\nonumber\\
&
- 6144(20\gamma^6 + 42\gamma^4 + 18\gamma^2 + 1)\gamma^{12}\eta^2\alpha_{\Delta}^{10}
\nonumber\\
&
+ 2304(164\gamma^8 + 604\gamma^6 + 618\gamma^4 + 148\gamma^2 + 5)\gamma^8\eta^4\alpha_{\Delta}^8
\Big) (4\gamma^2\alpha_{\Delta}^2 + 9\eta^2)^4
\nonumber\\
&
+ 432\eta^2 M(x)^2 s'(x)^2 \Big(
65536(241\gamma^8 + 392\gamma^6 + 168\gamma^4 + 8\gamma^2 + 1)\gamma^{20}\alpha_{\Delta}^{16}
\nonumber\\
&
- 98304(605\gamma^{10} + 2762\gamma^8 + 3122\gamma^6 + 1087\gamma^4 + 37\gamma^2 + 1)
\gamma^{16}\eta^2\alpha_{\Delta}^{14}
\nonumber\\
&
+ 27\eta^4\Big[
81\eta^4\Big(
9\eta^2\big[
-648\gamma^2(18\gamma^2 + 19)\eta^4\alpha_{\Delta}^2
\nonumber\\
&
+ 144\gamma^4(526\gamma^4 + 1152\gamma^2 + 623)\eta^2\alpha_{\Delta}^4
\nonumber\\
&
- 128\gamma^6(951\gamma^6 + 2844\gamma^4 + 2519\gamma^2 + 586)\alpha_{\Delta}^6 + 729\eta^6
\big]
\nonumber\\
&
+ 256\gamma^8(2934\gamma^8 + 11868\gamma^6 + 14852\gamma^4 + 6236\gamma^2 + 623)\alpha_{\Delta}^8
\Big)
\nonumber\\
&
- 18432\gamma^{10}\left(
2(521\gamma^8 + 3066\gamma^6 + 5275\gamma^4 + 3202\gamma^2 + 603)\gamma^2 + 19
\right)\eta^2\alpha_{\Delta}^{10}
\nonumber\\
&
+ 4096\gamma^{12}(563\gamma^{12} + 6192\gamma^{10} + 15494\gamma^8
+ 12852\gamma^6 + 3489\gamma^4 + 100\gamma^2 + 1)\alpha_{\Delta}^{12}
\Big)
\Bigg)
\end{align}

{\underline{Symmetric bounce:}}

\begin{align}
    \mathcal{C} &= 36 \left(\left(16 \alpha_\Delta^4+27 z^4+48 \alpha_\Delta^2 z^2\right) M'(x)-48 \alpha_\Delta^2 z M(x) s'(x)\right) \\
    \mathcal{D} &= 432 \Bigl(
\,384 \alpha_\Delta^{2} z\, M(x)
\bigl(-8 \alpha_\Delta^{4} + 27 z^{4} - 6 \alpha_\Delta^{2} z^{2}\bigr)
M'(x)\, s'(x)
\nonumber\\
&+ \bigl(4 \alpha_\Delta^{2} + 9 z^{2}\bigr)^{2}
\bigl(16 \alpha_\Delta^{4} + 45 z^{4} - 24 \alpha_\Delta^{2} z^{2}\bigr)
M'(x)^{2}
\nonumber\\
&+ 432 z^{2} M(x)^{2}
\bigl(160 \alpha_\Delta^{4} + 27 z^{4} - 96 \alpha_\Delta^{2} z^{2}\bigr)
s'(x)^{2}
\Bigr)
\end{align}

\newpage

\bibliographystyle{unsrtnat}
\bibliography{ref.bib}

\begin{thebibliography}{80}
\providecommand{\natexlab}[1]{#1}
\providecommand{\url}[1]{\texttt{#1}}
\expandafter\ifx\csname urlstyle\endcsname\relax
  \providecommand{\doi}[1]{doi: #1}\else
  \providecommand{\doi}{doi: \begingroup \urlstyle{rm}\Url}\fi

\bibitem[Ashtekar and Bojowald(2006)]{Ashtekar:2005qt}
Abhay Ashtekar and Martin Bojowald.
\newblock {Quantum geometry and the Schwarzschild singularity}.
\newblock \emph{Class. Quant. Grav.}, 23:\penalty0 391--411, 2006.
\newblock \doi{10.1088/0264-9381/23/2/008}.

\bibitem[Modesto(2006)]{Modesto:2005zm}
Leonardo Modesto.
\newblock {Loop quantum black hole}.
\newblock \emph{Class. Quant. Grav.}, 23:\penalty0 5587--5602, 2006.
\newblock \doi{10.1088/0264-9381/23/18/006}.

\bibitem[Boehmer and Vandersloot(2007)]{Boehmer:2007ket}
Christian~G. Boehmer and Kevin Vandersloot.
\newblock {Loop Quantum Dynamics of the Schwarzschild Interior}.
\newblock \emph{Phys. Rev. D}, 76:\penalty0 104030, 2007.
\newblock \doi{10.1103/PhysRevD.76.104030}.

\bibitem[Chiou et~al.(2012)Chiou, Ni, and Tang]{Chiou:2012pg}
Dah-Wei Chiou, Wei-Tou Ni, and Alf Tang.
\newblock {Loop quantization of spherically symmetric midisuperspaces and loop quantum geometry of the maximally extended Schwarzschild spacetime}.
\newblock 12 2012.

\bibitem[Gambini et~al.(2014)Gambini, Olmedo, and Pullin]{Gambini:2013hna}
Rodolfo Gambini, Javier Olmedo, and Jorge Pullin.
\newblock {Quantum black holes in Loop Quantum Gravity}.
\newblock \emph{Class. Quant. Grav.}, 31:\penalty0 095009, 2014.
\newblock \doi{10.1088/0264-9381/31/9/095009}.

\bibitem[Brahma(2015)]{Brahma:2014gca}
Suddhasattwa Brahma.
\newblock {Spherically symmetric canonical quantum gravity}.
\newblock \emph{Phys. Rev. D}, 91\penalty0 (12):\penalty0 124003, 2015.
\newblock \doi{10.1103/PhysRevD.91.124003}.

\bibitem[Dadhich et~al.(2015)Dadhich, Joe, and Singh]{Dadhich:2015ora}
Naresh Dadhich, Anton Joe, and Parampreet Singh.
\newblock {Emergence of the product of constant curvature spaces in loop quantum cosmology}.
\newblock \emph{Class. Quant. Grav.}, 32\penalty0 (18):\penalty0 185006, 2015.
\newblock \doi{10.1088/0264-9381/32/18/185006}.

\bibitem[Tibrewala(2014)]{Tibrewala:2013kba}
Rakesh Tibrewala.
\newblock {Inhomogeneities, loop quantum gravity corrections, constraint algebra and general covariance}.
\newblock \emph{Class. Quant. Grav.}, 31:\penalty0 055010, 2014.
\newblock \doi{10.1088/0264-9381/31/5/055010}.

\bibitem[Ben~Achour et~al.(2018{\natexlab{a}})Ben~Achour, Lamy, Liu, and Noui]{BenAchour:2017ivq}
Jibril Ben~Achour, Frederic Lamy, Hongguang Liu, and Karim Noui.
\newblock {Non-singular black holes and the Limiting Curvature Mechanism: A Hamiltonian perspective}.
\newblock \emph{JCAP}, 05:\penalty0 072, 2018{\natexlab{a}}.
\newblock \doi{10.1088/1475-7516/2018/05/072}.

\bibitem[Yonika et~al.(2018)Yonika, Khanna, and Singh]{Yonika:2017qgo}
Alec Yonika, Gaurav Khanna, and Parampreet Singh.
\newblock {Von-Neumann Stability and Singularity Resolution in Loop Quantized Schwarzschild Black Hole}.
\newblock \emph{Class. Quant. Grav.}, 35\penalty0 (4):\penalty0 045007, 2018.
\newblock \doi{10.1088/1361-6382/aaa18d}.

\bibitem[D'Ambrosio et~al.(2021)D'Ambrosio, Christodoulou, Martin-Dussaud, Rovelli, and Soltani]{DAmbrosio:2020mut}
Fabio D'Ambrosio, Marios Christodoulou, Pierre Martin-Dussaud, Carlo Rovelli, and Farshid Soltani.
\newblock {End of a black hole{\textquoteright}s evaporation}.
\newblock \emph{Phys. Rev. D}, 103\penalty0 (10):\penalty0 106014, 2021.
\newblock \doi{10.1103/PhysRevD.103.106014}.

\bibitem[Olmedo et~al.(2017)Olmedo, Saini, and Singh]{Olmedo:2017lvt}
Javier Olmedo, Sahil Saini, and Parampreet Singh.
\newblock {From black holes to white holes: a quantum gravitational, symmetric bounce}.
\newblock \emph{Class. Quant. Grav.}, 34\penalty0 (22):\penalty0 225011, 2017.
\newblock \doi{10.1088/1361-6382/aa8da8}.

\bibitem[Ashtekar et~al.(2018{\natexlab{a}})Ashtekar, Olmedo, and Singh]{Ashtekar:2018lag}
Abhay Ashtekar, Javier Olmedo, and Parampreet Singh.
\newblock {Quantum Transfiguration of Kruskal Black Holes}.
\newblock \emph{Phys. Rev. Lett.}, 121\penalty0 (24):\penalty0 241301, 2018{\natexlab{a}}.
\newblock \doi{10.1103/PhysRevLett.121.241301}.

\bibitem[Ashtekar et~al.(2018{\natexlab{b}})Ashtekar, Olmedo, and Singh]{Ashtekar:2018cay}
Abhay Ashtekar, Javier Olmedo, and Parampreet Singh.
\newblock {Quantum extension of the Kruskal spacetime}.
\newblock \emph{Phys. Rev. D}, 98\penalty0 (12):\penalty0 126003, 2018{\natexlab{b}}.
\newblock \doi{10.1103/PhysRevD.98.126003}.

\bibitem[Bojowald et~al.(2018)Bojowald, Brahma, and Yeom]{Bojowald:2018xxu}
Martin Bojowald, Suddhasattwa Brahma, and Dong-han Yeom.
\newblock {Effective line elements and black-hole models in canonical loop quantum gravity}.
\newblock \emph{Phys. Rev. D}, 98\penalty0 (4):\penalty0 046015, 2018.
\newblock \doi{10.1103/PhysRevD.98.046015}.

\bibitem[Ben~Achour et~al.(2018{\natexlab{b}})Ben~Achour, Lamy, Liu, and Noui]{BenAchour:2018khr}
Jibril Ben~Achour, Fr{\'e}d{\'e}ric Lamy, Hongguang Liu, and Karim Noui.
\newblock {Polymer Schwarzschild black hole: An effective metric}.
\newblock \emph{EPL}, 123\penalty0 (2):\penalty0 20006, 2018{\natexlab{b}}.
\newblock \doi{10.1209/0295-5075/123/20006}.

\bibitem[Bodendorfer et~al.(2019)Bodendorfer, Mele, and M{\"u}nch]{Bodendorfer:2019cyv}
Norbert Bodendorfer, Fabio~M. Mele, and Johannes M{\"u}nch.
\newblock {Effective Quantum Extended Spacetime of Polymer Schwarzschild Black Hole}.
\newblock \emph{Class. Quant. Grav.}, 36\penalty0 (19):\penalty0 195015, 2019.
\newblock \doi{10.1088/1361-6382/ab3f16}.

\bibitem[Alesci et~al.(2019)Alesci, Bahrami, and Pranzetti]{Alesci:2019pbs}
Emanuele Alesci, Sina Bahrami, and Daniele Pranzetti.
\newblock {Quantum gravity predictions for black hole interior geometry}.
\newblock \emph{Phys. Lett. B}, 797:\penalty0 134908, 2019.
\newblock \doi{10.1016/j.physletb.2019.134908}.

\bibitem[Assanioussi et~al.(2020)Assanioussi, Dapor, and Liegener]{Assanioussi:2019twp}
Mehdi Assanioussi, Andrea Dapor, and Klaus Liegener.
\newblock {Perspectives on the dynamics in a loop quantum gravity effective description of black hole interiors}.
\newblock \emph{Phys. Rev. D}, 101\penalty0 (2):\penalty0 026002, 2020.
\newblock \doi{10.1103/PhysRevD.101.026002}.

\bibitem[Benitez et~al.(2020)Benitez, Gambini, Lehner, Liebling, and Pullin]{Benitez:2020szx}
Florencia Benitez, Rodolfo Gambini, Luis Lehner, Steve Liebling, and Jorge Pullin.
\newblock {Critical collapse of a scalar field in semiclassical loop quantum gravity}.
\newblock \emph{Phys. Rev. Lett.}, 124\penalty0 (7):\penalty0 071301, 2020.
\newblock \doi{10.1103/PhysRevLett.124.071301}.

\bibitem[Gan et~al.(2020)Gan, Santos, Shu, and Wang]{Gan:2020dkb}
Wen-Cong Gan, Nilton~O. Santos, Fu-Wen Shu, and Anzhong Wang.
\newblock {Properties of the spherically symmetric polymer black holes}.
\newblock \emph{Phys. Rev. D}, 102:\penalty0 124030, 2020.
\newblock \doi{10.1103/PhysRevD.102.124030}.

\bibitem[Gambini et~al.(2021)Gambini, Olmedo, and Pullin]{Gambini:2020qhx}
Rodolfo Gambini, Javier Olmedo, and Jorge Pullin.
\newblock {Loop Quantum Black Hole Extensions Within the Improved Dynamics}.
\newblock \emph{Front. Astron. Space Sci.}, 8:\penalty0 74, 2021.
\newblock \doi{10.3389/fspas.2021.647241}.

\bibitem[Husain et~al.(2022{\natexlab{a}})Husain, Kelly, Santacruz, and Wilson-Ewing]{Husain:2021ojz}
Viqar Husain, Jarod~George Kelly, Robert Santacruz, and Edward Wilson-Ewing.
\newblock {Quantum Gravity of Dust Collapse: Shock Waves from Black Holes}.
\newblock \emph{Phys. Rev. Lett.}, 128\penalty0 (12):\penalty0 121301, 2022{\natexlab{a}}.
\newblock \doi{10.1103/PhysRevLett.128.121301}.

\bibitem[Husain et~al.(2022{\natexlab{b}})Husain, Kelly, Santacruz, and Wilson-Ewing]{Husain:2022gwp}
Viqar Husain, Jarod~George Kelly, Robert Santacruz, and Edward Wilson-Ewing.
\newblock {Fate of quantum black holes}.
\newblock \emph{Phys. Rev. D}, 106\penalty0 (2):\penalty0 024014, 2022{\natexlab{b}}.
\newblock \doi{10.1103/PhysRevD.106.024014}.

\bibitem[Li and Singh(2021)]{Li:2021snn}
Bao-Fei Li and Parampreet Singh.
\newblock {Does the Loop Quantum $\mu_o$ Scheme Permit Black Hole Formation?}
\newblock \emph{Universe}, 7\penalty0 (11):\penalty0 406, 2021.
\newblock \doi{10.3390/universe7110406}.

\bibitem[Gan et~al.(2022)Gan, Ongole, Alesci, An, Shu, and Wang]{Gan:2022mle}
Wen-Cong Gan, Geeth Ongole, Emanuele Alesci, Yang An, Fu-Wen Shu, and Anzhong Wang.
\newblock {Understanding quantum black holes from quantum reduced loop gravity}.
\newblock \emph{Phys. Rev. D}, 106\penalty0 (12):\penalty0 126013, 2022.
\newblock \doi{10.1103/PhysRevD.106.126013}.

\bibitem[Kelly et~al.(2020)Kelly, Santacruz, and Wilson-Ewing]{Kelly:2020uwj}
Jarod~George Kelly, Robert Santacruz, and Edward Wilson-Ewing.
\newblock {Effective loop quantum gravity framework for vacuum spherically symmetric spacetimes}.
\newblock \emph{Phys. Rev. D}, 102\penalty0 (10):\penalty0 106024, 2020.
\newblock \doi{10.1103/PhysRevD.102.106024}.

\bibitem[Gambini et~al.(2020)Gambini, Olmedo, and Pullin]{Gambini:2020nsf}
R.~Gambini, J.~Olmedo, and J.~Pullin.
\newblock {Spherically symmetric loop quantum gravity: analysis of improved dynamics}.
\newblock \emph{Class. Quant. Grav.}, 37\penalty0 (20):\penalty0 205012, 2020.
\newblock \doi{10.1088/1361-6382/aba842}.

\bibitem[Han and Liu(2022)]{Han:2020uhb}
Muxin Han and Hongguang Liu.
\newblock {Improved effective dynamics of loop-quantum-gravity black hole and Nariai limit}.
\newblock \emph{Class. Quant. Grav.}, 39\penalty0 (3):\penalty0 035011, 2022.
\newblock \doi{10.1088/1361-6382/ac44a0}.

\bibitem[Zhang(2021)]{Zhang:2021xoa}
Cong Zhang.
\newblock {Reduced phase space quantization of black holes: Path integrals and effective dynamics}.
\newblock \emph{Phys. Rev. D}, 104\penalty0 (12):\penalty0 126003, 2021.
\newblock \doi{10.1103/PhysRevD.104.126003}.

\bibitem[M{\"u}nch et~al.(2023)M{\"u}nch, Perez, Speziale, and Viollet]{Munch:2022teq}
Johannes M{\"u}nch, Alejandro Perez, Simone Speziale, and Sami Viollet.
\newblock {Generic features of a polymer quantum black hole}.
\newblock \emph{Class. Quant. Grav.}, 40\penalty0 (13):\penalty0 135003, 2023.
\newblock \doi{10.1088/1361-6382/accccd}.

\bibitem[Lewandowski et~al.(2023)Lewandowski, Ma, Yang, and Zhang]{Lewandowski:2022zce}
Jerzy Lewandowski, Yongge Ma, Jinsong Yang, and Cong Zhang.
\newblock {Quantum Oppenheimer-Snyder and Swiss Cheese Models}.
\newblock \emph{Phys. Rev. Lett.}, 130\penalty0 (10):\penalty0 101501, 2023.
\newblock \doi{10.1103/PhysRevLett.130.101501}.

\bibitem[Giesel et~al.(2021)Giesel, Li, and Singh]{Giesel:2021dug}
Kristina Giesel, Bao-Fei Li, and Parampreet Singh.
\newblock {Nonsingular quantum gravitational dynamics of an Lema{\^\i}tre-Tolman-Bondi dust shell model: The role of quantization prescriptions}.
\newblock \emph{Phys. Rev. D}, 104\penalty0 (10):\penalty0 106017, 2021.
\newblock \doi{10.1103/PhysRevD.104.106017}.

\bibitem[Giesel et~al.(2023)Giesel, Han, Li, Liu, and Singh]{Giesel:2022rxi}
Kristina Giesel, Muxin Han, Bao-Fei Li, Hongguang Liu, and Parampreet Singh.
\newblock {Spherical symmetric gravitational collapse of a dust cloud: Polymerized dynamics in reduced phase space}.
\newblock \emph{Phys. Rev. D}, 107\penalty0 (4):\penalty0 044047, 2023.
\newblock \doi{10.1103/PhysRevD.107.044047}.

\bibitem[Han and Liu(2024)]{Han:2022rsx}
Muxin Han and Hongguang Liu.
\newblock {Covariant {\ensuremath{\mu}}{\textasciimacron}-scheme effective dynamics, mimetic gravity, and nonsingular black holes: Applications to spherically symmetric quantum gravity}.
\newblock \emph{Phys. Rev. D}, 109\penalty0 (8):\penalty0 084033, 2024.
\newblock \doi{10.1103/PhysRevD.109.084033}.

\bibitem[Giesel et~al.(2024{\natexlab{a}})Giesel, Liu, Rullit, Singh, and Weigl]{Giesel:2023tsj}
Kristina Giesel, Hongguang Liu, Eric Rullit, Parampreet Singh, and Stefan~Andreas Weigl.
\newblock {Embedding generalized Lema{\^\i}tre-Tolman-Bondi models in polymerized spherically symmetric spacetimes}.
\newblock \emph{Phys. Rev. D}, 110\penalty0 (10):\penalty0 104017, 2024{\natexlab{a}}.
\newblock \doi{10.1103/PhysRevD.110.104017}.

\bibitem[Giesel et~al.(2025)Giesel, Liu, Singh, and Weigl]{Giesel:2024mps}
Kristina Giesel, Hongguang Liu, Parampreet Singh, and Stefan~Andreas Weigl.
\newblock {Regular black holes and their relationship to polymerized models and mimetic gravity}.
\newblock \emph{Phys. Rev. D}, 111\penalty0 (6):\penalty0 064064, 2025.
\newblock \doi{10.1103/PhysRevD.111.064064}.

\bibitem[Cafaro and Lewandowski(2024)]{Cafaro:2024vrw}
Luca Cafaro and Jerzy Lewandowski.
\newblock {Status of Birkhoff{\textquoteright}s theorem in the polymerized semiclassical regime of loop quantum gravity}.
\newblock \emph{Phys. Rev. D}, 110\penalty0 (2):\penalty0 024072, 2024.
\newblock \doi{10.1103/PhysRevD.110.024072}.

\bibitem[Giesel and Liu(2025)]{Giesel:2025kdl}
Kristina Giesel and Hongguang Liu.
\newblock {From Principles to Effective Models: A Constructive Framework for Effective Covariant Actions with a Unique Vacuum Solution}.
\newblock 12 2025.

\bibitem[Bojowald et~al.(2008)Bojowald, Harada, and Tibrewala]{Bojowald:2008ja}
Martin Bojowald, Tomohiro Harada, and Rakesh Tibrewala.
\newblock {Lemaitre-Tolman-Bondi collapse from the perspective of loop quantum gravity}.
\newblock \emph{Phys. Rev. D}, 78:\penalty0 064057, 2008.
\newblock \doi{10.1103/PhysRevD.78.064057}.

\bibitem[Bojowald et~al.(2009)Bojowald, Reyes, and Tibrewala]{Bojowald:2009ih}
Martin Bojowald, Juan~D. Reyes, and Rakesh Tibrewala.
\newblock {Non-marginal LTB-like models with inverse triad corrections from loop quantum gravity}.
\newblock \emph{Phys. Rev. D}, 80:\penalty0 084002, 2009.
\newblock \doi{10.1103/PhysRevD.80.084002}.

\bibitem[M{\"u}nch(2021)]{Munch:2020czs}
Johannes M{\"u}nch.
\newblock {Effective quantum dust collapse via surface matching}.
\newblock \emph{Class. Quant. Grav.}, 38\penalty0 (17):\penalty0 175015, 2021.
\newblock \doi{10.1088/1361-6382/ac103e}.

\bibitem[Fazzini et~al.(2023)Fazzini, Rovelli, and Soltani]{Fazzini:2023scu}
Francesco Fazzini, Carlo Rovelli, and Farshid Soltani.
\newblock {Painlev{\'e}-Gullstrand coordinates discontinuity in the quantum Oppenheimer-Snyder model}.
\newblock \emph{Phys. Rev. D}, 108\penalty0 (4):\penalty0 044009, 2023.
\newblock \doi{10.1103/PhysRevD.108.044009}.

\bibitem[Han et~al.(2023)Han, Rovelli, and Soltani]{Han:2023wxg}
Muxin Han, Carlo Rovelli, and Farshid Soltani.
\newblock {Geometry of the black-to-white hole transition within a single asymptotic region}.
\newblock \emph{Phys. Rev. D}, 107\penalty0 (6):\penalty0 064011, 2023.
\newblock \doi{10.1103/PhysRevD.107.064011}.

\bibitem[Fazzini et~al.(2024)Fazzini, Husain, and Wilson-Ewing]{Fazzini:2023ova}
Francesco Fazzini, Viqar Husain, and Edward Wilson-Ewing.
\newblock {Shell-crossings and shock formation during gravitational collapse in effective loop quantum gravity}.
\newblock \emph{Phys. Rev. D}, 109\penalty0 (8):\penalty0 084052, 2024.
\newblock \doi{10.1103/PhysRevD.109.084052}.

\bibitem[Cipriani et~al.(2024)Cipriani, Fazzini, and Wilson-Ewing]{Cipriani:2024nhx}
Lorenzo Cipriani, Francesco Fazzini, and Edward Wilson-Ewing.
\newblock {Gravitational collapse in effective loop quantum gravity: Beyond marginally bound configurations}.
\newblock \emph{Phys. Rev. D}, 110\penalty0 (6):\penalty0 066004, 2024.
\newblock \doi{10.1103/PhysRevD.110.066004}.

\bibitem[Han et~al.(2025)Han, Liu, Qu, Vidotto, and Zhang]{Han:2024ydv}
Muxin Han, Hongguang Liu, Dongxue Qu, Francesca Vidotto, and Cong Zhang.
\newblock {Cosmological dynamics from covariant loop quantum gravity with scalar matter}.
\newblock \emph{Phys. Rev. D}, 111\penalty0 (8):\penalty0 086012, 2025.
\newblock \doi{10.1103/PhysRevD.111.086012}.

\bibitem[Ou and Zhang(2025)]{Ou:2025bbv}
Minyan Ou and Xiangdong Zhang.
\newblock {Quantum Oppenheimer-Snyder models in loop quantum cosmology with Lorentz term}.
\newblock \emph{Phys. Rev. D}, 112\penalty0 (12):\penalty0 126016, 2025.
\newblock \doi{10.1103/d4fp-nwbf}.

\bibitem[Lasky et~al.(2006)Lasky, Lun, and Burston]{Lasky:2006hq}
Paul~D. Lasky, Anthony W.~C. Lun, and Raymond~B. Burston.
\newblock {Initial value formalism for dust collapse}.
\newblock 6 2006.

\bibitem[Fazzini and Mehmood(2025)]{Fazzini:2025zrq}
Francesco Fazzini and Hassan Mehmood.
\newblock {Weak solutions in Einstein theory and beyond}.
\newblock \emph{Phys. Rev. D}, 112\penalty0 (6):\penalty0 064064, 2025.
\newblock \doi{10.1103/6mrq-f9qc}.

\bibitem[Liu and Qu(2025)]{Liu:2025fil}
Hongguang Liu and Dongxue Qu.
\newblock {Quantum induced shock dynamics in gravitational collapse: insights from effective models and numerical frameworks}.
\newblock 4 2025.

\bibitem[Sahlmann and Zhang(2025)]{Sahlmann:2025fde}
Hanno Sahlmann and Cong Zhang.
\newblock {Dust shell in effective loop quantum black hole model}.
\newblock \emph{Phys. Rev. D}, 112\penalty0 (8):\penalty0 084079, 2025.
\newblock \doi{10.1103/r33z-17wy}.

\bibitem[Fazzini(2026)]{Fazzini:2025nse}
Francesco Fazzini.
\newblock {Effective LQG dynamics of a thin shell and the fate of a collapsing star}.
\newblock \emph{Phys. Rev. D}, 113\penalty0 (2):\penalty0 026020, 2026.
\newblock \doi{10.1103/hj85-317t}.

\bibitem[Yang et~al.(2009)Yang, Ding, and Ma]{Yang:2009fp}
Jinsong Yang, You Ding, and Yongge Ma.
\newblock {Alternative quantization of the Hamiltonian in loop quantum cosmology II: Including the Lorentz term}.
\newblock \emph{Phys. Lett. B}, 682:\penalty0 1--7, 2009.
\newblock \doi{10.1016/j.physletb.2009.10.072}.

\bibitem[Li et~al.(2018)Li, Singh, and Wang]{Li:2018opr}
Bao-Fei Li, Parampreet Singh, and Anzhong Wang.
\newblock {Towards Cosmological Dynamics from Loop Quantum Gravity}.
\newblock \emph{Phys. Rev. D}, 97\penalty0 (8):\penalty0 084029, 2018.
\newblock \doi{10.1103/PhysRevD.97.084029}.

\bibitem[Dapor and Liegener(2018)]{Dapor:2017rwv}
Andrea Dapor and Klaus Liegener.
\newblock {Cosmological Effective Hamiltonian from full Loop Quantum Gravity Dynamics}.
\newblock \emph{Phys. Lett. B}, 785:\penalty0 506--510, 2018.
\newblock \doi{10.1016/j.physletb.2018.09.005}.

\bibitem[Han and Liu(2020{\natexlab{a}})]{Han:2019vpw}
Muxin Han and Hongguang Liu.
\newblock {Effective Dynamics from Coherent State Path Integral of Full Loop Quantum Gravity}.
\newblock \emph{Phys. Rev. D}, 101\penalty0 (4):\penalty0 046003, 2020{\natexlab{a}}.
\newblock \doi{10.1103/PhysRevD.101.046003}.

\bibitem[Han and Liu(2020{\natexlab{b}})]{Han:2019feb}
Muxin Han and Hongguang Liu.
\newblock {Improved $\overline{\mu}$-scheme effective dynamics of full loop quantum gravity}.
\newblock \emph{Phys. Rev. D}, 102\penalty0 (6):\penalty0 064061, 2020{\natexlab{b}}.
\newblock \doi{10.1103/PhysRevD.102.064061}.

\bibitem[Han and Liu(2021)]{Han:2021cwb}
Muxin Han and Hongguang Liu.
\newblock {Loop quantum gravity on dynamical lattice and improved cosmological effective dynamics with inflaton}.
\newblock \emph{Phys. Rev. D}, 104\penalty0 (2):\penalty0 024011, 2021.
\newblock \doi{10.1103/PhysRevD.104.024011}.

\bibitem[Giesel and Liu(2023)]{Giesel:2023euq}
Kristina Giesel and Hongguang Liu.
\newblock {Dynamically implementing the $\overline{\mu}$-scheme in cosmological and spherically symmetric models in an extended phase space model}.
\newblock \emph{Universe}, 9\penalty0 (4):\penalty0 176, 2023.
\newblock \doi{10.3390/universe9040176}.

\bibitem[Giesel et~al.(2024{\natexlab{b}})Giesel, Liu, Singh, and Weigl]{Giesel:2023hys}
Kristina Giesel, Hongguang Liu, Parampreet Singh, and Stefan~Andreas Weigl.
\newblock {Generalized analysis of a dust collapse in effective loop quantum gravity: Fate of shocks and covariance}.
\newblock \emph{Phys. Rev. D}, 110\penalty0 (10):\penalty0 104016, 2024{\natexlab{b}}.
\newblock \doi{10.1103/PhysRevD.110.104016}.

\bibitem[Lemaitre(1933)]{Lemaitre:1933gd}
G.~Lemaitre.
\newblock {The expanding universe}.
\newblock \emph{Annales Soc. Sci. Bruxelles A}, 53:\penalty0 51--85, 1933.
\newblock \doi{10.1023/A:1018855621348}.

\bibitem[Tolman(1934)]{Tolman:1934za}
Richard~C. Tolman.
\newblock {Effect of imhomogeneity on cosmological models}.
\newblock \emph{Proc. Nat. Acad. Sci.}, 20:\penalty0 169--176, 1934.
\newblock \doi{10.1073/pnas.20.3.169}.

\bibitem[Bondi(1947)]{Bondi:1947fta}
H.~Bondi.
\newblock {Spherically symmetrical models in general relativity}.
\newblock \emph{Mon. Not. Roy. Astron. Soc.}, 107:\penalty0 410--425, 1947.
\newblock \doi{10.1093/mnras/107.5-6.410}.

\bibitem[Rovelli and Smolin(1995)]{Rovelli:1994ge}
Carlo Rovelli and Lee Smolin.
\newblock {Discreteness of area and volume in quantum gravity}.
\newblock \emph{Nucl. Phys. B}, 442:\penalty0 593--622, 1995.
\newblock \doi{10.1016/0550-3213(95)00150-Q}.
\newblock [Erratum: Nucl.Phys.B 456, 753--754 (1995)].

\bibitem[Ashtekar and Lewandowski(1997)]{Ashtekar:1996eg}
Abhay Ashtekar and Jerzy Lewandowski.
\newblock {Quantum theory of geometry. 1: Area operators}.
\newblock \emph{Class. Quant. Grav.}, 14:\penalty0 A55--A82, 1997.
\newblock \doi{10.1088/0264-9381/14/1A/006}.

\bibitem[Giesel and Thiemann(2007{\natexlab{a}})]{Giesel:2006uj}
K.~Giesel and T.~Thiemann.
\newblock {Algebraic Quantum Gravity (AQG). I. Conceptual Setup}.
\newblock \emph{Class. Quant. Grav.}, 24:\penalty0 2465--2498, 2007{\natexlab{a}}.
\newblock \doi{10.1088/0264-9381/24/10/003}.

\bibitem[Thiemann(2001)]{Thiemann:2000bw}
Thomas Thiemann.
\newblock {Gauge field theory coherent states (GCS): 1. General properties}.
\newblock \emph{Class. Quant. Grav.}, 18:\penalty0 2025--2064, 2001.
\newblock \doi{10.1088/0264-9381/18/11/304}.

\bibitem[Thiemann and Winkler(2001{\natexlab{a}})]{Thiemann:2000ca}
T.~Thiemann and O.~Winkler.
\newblock {Gauge field theory coherent states (GCS). 2. Peakedness properties}.
\newblock \emph{Class. Quant. Grav.}, 18:\penalty0 2561--2636, 2001{\natexlab{a}}.
\newblock \doi{10.1088/0264-9381/18/14/301}.

\bibitem[Thiemann and Winkler(2001{\natexlab{b}})]{Thiemann:2000bx}
T.~Thiemann and O.~Winkler.
\newblock {Gauge field theory coherent states (GCS): 3. Ehrenfest theorems}.
\newblock \emph{Class. Quant. Grav.}, 18:\penalty0 4629--4682, 2001{\natexlab{b}}.
\newblock \doi{10.1088/0264-9381/18/21/315}.

\bibitem[Thiemann(2006)]{Thiemann:2002vj}
Thomas Thiemann.
\newblock {Complexifier coherent states for quantum general relativity}.
\newblock \emph{Class. Quant. Grav.}, 23:\penalty0 2063--2118, 2006.
\newblock \doi{10.1088/0264-9381/23/6/013}.

\bibitem[Giesel and Thiemann(2007{\natexlab{b}})]{Giesel:2006um}
K.~Giesel and T.~Thiemann.
\newblock {Algebraic quantum gravity (AQG). III. Semiclassical perturbation theory}.
\newblock \emph{Class. Quant. Grav.}, 24:\penalty0 2565--2588, 2007{\natexlab{b}}.
\newblock \doi{10.1088/0264-9381/24/10/005}.

\bibitem[Ashtekar et~al.(1998)Ashtekar, Baez, Corichi, and Krasnov]{Ashtekar:1997yu}
A.~Ashtekar, J.~Baez, A.~Corichi, and Kirill Krasnov.
\newblock {Quantum geometry and black hole entropy}.
\newblock \emph{Phys. Rev. Lett.}, 80:\penalty0 904--907, 1998.
\newblock \doi{10.1103/PhysRevLett.80.904}.

\bibitem[Assanioussi et~al.(2018)Assanioussi, Dapor, Liegener, and Paw{\l}owski]{Assanioussi:2018hee}
Mehdi Assanioussi, Andrea Dapor, Klaus Liegener, and Tomasz Paw{\l}owski.
\newblock {Emergent de Sitter Epoch of the Quantum Cosmos from Loop Quantum Cosmology}.
\newblock \emph{Phys. Rev. Lett.}, 121\penalty0 (8):\penalty0 081303, 2018.
\newblock \doi{10.1103/PhysRevLett.121.081303}.

\bibitem[Fazzini et~al.()Fazzini, Giesel, and Rullit]{tba}
Francesco Fazzini, Kristina Giesel, and Eric Rullit.
\newblock {Shell-crossing singularity formation in different models of quantum gravitational collapse}.
\newblock to appear.

\bibitem[Assanioussi et~al.(2019)Assanioussi, Dapor, Liegener, and Paw{\l}owski]{Assanioussi:2019iye}
Mehdi Assanioussi, Andrea Dapor, Klaus Liegener, and Tomasz Paw{\l}owski.
\newblock {Emergent de Sitter epoch of the Loop Quantum Cosmos: a detailed analysis}.
\newblock \emph{Phys. Rev. D}, 100\penalty0 (8):\penalty0 084003, 2019.
\newblock \doi{10.1103/PhysRevD.100.084003}.

\bibitem[Ashtekar et~al.(2006)Ashtekar, Pawlowski, and Singh]{Ashtekar:2006wn}
Abhay Ashtekar, Tomasz Pawlowski, and Parampreet Singh.
\newblock {Quantum Nature of the Big Bang: Improved dynamics}.
\newblock \emph{Phys. Rev. D}, 74:\penalty0 084003, 2006.
\newblock \doi{10.1103/PhysRevD.74.084003}.

\bibitem[Hayward(1996)]{Hayward:1994bu}
Sean~A. Hayward.
\newblock {Gravitational energy in spherical symmetry}.
\newblock \emph{Phys. Rev. D}, 53:\penalty0 1938--1949, 1996.
\newblock \doi{10.1103/PhysRevD.53.1938}.

\bibitem[Bardeen(1968)]{Bardeen68}
James~M Bardeen.
\newblock Non-singular general-relativistic gravitational collapse, in conference proceedings of gr5, 1968.

\bibitem[Kiefer and Schmitz(2019)]{Kiefer:2019csi}
Claus Kiefer and Tim Schmitz.
\newblock {Singularity avoidance for collapsing quantum dust in the Lema{\^\i}tre-Tolman-Bondi model}.
\newblock \emph{Phys. Rev. D}, 99\penalty0 (12):\penalty0 126010, 2019.
\newblock \doi{10.1103/PhysRevD.99.126010}.

\end{thebibliography}

\end{document}